\documentstyle[aps,floats,epsf,twocolumn]{revtex}

\newcommand{\bk}{{\bf k}}

\newcommand{\br}{{\bf r}}

\newcommand{\bA}{{\bf A}}
\newcommand{\bv}{{\bf v}}
\newcommand{\bB}{{\bf B}}

\begin{document} \draft

\title{Quasiparticles and Vortices in Unconventional Superconductors}

\author{O. Vafek, A. Melikyan, M. Franz$^*$ and Z. Te\v{s}anovi\'c}
\address{Department of Physics and Astronomy, Johns Hopkins University,
Baltimore, MD 21218
\\ {\rm(\today)}
}
%
\address{~
\parbox{14cm}{\rm
\medskip 
Quasiparticles in the vortex lattice
of strongly type-II superconductors are investigated by means of
a singular gauge transformation applied to the tight binding lattice 
Bogoliubov-de Gennes Hamiltonian.  
We present a detailed derivation of the gauge invariant
effective low energy 
Hamiltonian for the quasiparticle-vortex system
and show how the physics of the ``Doppler shift"
and ``Berry phase" can be incorporated at the Hamiltonian
level by working in the singular gauge. In particular, we 
show that the ``Berry phase" effect manifests itself in the
effective Hamiltonian through a half-flux Aharonov-Bohm scattering 
of quasiparticles off vortices and stress the important role that 
this effect plays in the quasiparticle dynamics.
Full numerical solutions in the regime
of intermediate fields $H_{c1} \ll B \ll H_{c2}$ are presented 
for model superconductors with $s$-, $p$- and 
$d$-wave symmetries and with square and triangular vortex lattices. 
For $s$- and $p$-wave cases we obtain low energy bound states in the core,
in agreement with the existing results. For $d$-wave case only extended
quasiparticle states exist. We investigate in detail the  nature of these 
extended states and provide 
comparison to the previous results within linearized ``Dirac fermion'' 
model. We also investigate internodal interference effects when 
vortex and ionic lattices have high degree of commensurability and 
discuss various specific choices for the singular gauge transformation.
}}

\maketitle

\section{Introduction}

In conventional ($s$-wave) superconductors the single particle fermionic
excitations (quasiparticles) are fully gapped everywhere on the Fermi
surface and the quasiparticle density of states vanishes below a specific
energy.  This has profound consequences for the traditional
phenomenology of superconductors. The gap in the fermionic spectrum leads
to the well known activated BCS form of the quasiparticle contribution to
various thermodynamic and transport properties. Furthermore, even as one
moves beyond the mean-field BCS theory, the absence of low-energy
quasiparticles in the superconducting state allows one to rewrite the
problem of superconducting fluctuations as a ``bosonic" theory, with the
role of bosons played by fluctuating Cooper pairs, after integrating out
``fermionic" degrees of freedom, i.e. the quasiparticles. In high
temperature superconductors (HTS), however, everything is different: the
cuprates appear to be accurately described by the $d_{x^2-y^2}$-wave order
parameter\cite{harlingen}, consequently allowing quasiparticle excitations
at arbitrary low energy near the nodal points.  These low energy fermionic
excitations appear to govern much of the thermodynamics and transport in
the HTS materials.  We are thus handed a new intellectual
challenge\cite{lee}: we must devise methods that can incorporate the low
energy fermionic excitations into the phenomenology of superconductors,
both within the mean-field BCS-like theory and beyond. 

This challenge is not trivial and has many diverse components:  low energy
quasiparticles are scattered by impurities in novel and unusual ways,
depending on the low energy density of states\cite{atkinson}; they
interact with external perturbations in ways not encountered in
conventional superconductors and these interactions give rise to new
phenomena\cite{Volovikdw93,sauls}; the low energy quasiparticles are 
expected to
qualitatively affect the quantum critical behavior of HTS.  Among many
aspects of this new quasiparticle phenomenology a particularly prominent
role is played by the low lying quasiparticle excitations in the mixed (or
vortex)  state. All HTS are extreme type-II systems and have a huge mixed
phase extending from the lower critical field $H_{c1}$ which is in the
range of 10-100 Gauss to the upper critical field $H_{c2}$ which can be as
large as 100-200 T. We suspect that in this large region the
interactions between quasiparticles and vortices play the essential role
in defining the nature of thermodynamic and transport properties. 

Such thermodynamic and transport properties are expected to be rather
different for distinct classes of unconventional superconductors.  This
difference stems from a complex motion of the quasiparticles under the
combined effects of both the magnetic field {\bf B} and the local drift
produced by chiral supercurrents of the vortex state. For example, in HTS
the $d_{x^2-y^2}$-wave nature of the gap function results in its vanishing
along nodal directions. Along these nodal directions
the pair-breaking induced by supercurrents has a particularly strong
effect. On the other hand, in unconventional superconductors with the
$p_{x \pm iy}$ pairing, Sr$_2$RuO$_4$ being a possible candidate\cite{Rice},
the spectrum is fully gapped but the order parameter is chiral even in the
absence of external magnetic field. This leads to two different types of
vortices for two different field orientations\cite{Volovikpw99,Matsumoto99}. 

Still, in
all these different situations, the quantum dynamics of quasiparticles in
the vortex state contains two essential common ingredients. {\em First},
there is a purely classical effect of a Doppler 
shift\cite{Volovikdw93,sauls}: 
a quasiparticle energy is shifted by a locally drifting superfluid, $E({\bf
k})\to E({\bf k}) - \hbar{\bf v}_s({\bf r})\cdot {\bf k}$, where ${\bf
v}_s({\bf r})$ is the local superfluid velocity.  ${\bf v}_s({\bf r})$
contains information about vortex configurations allowing us to connect
quasiparticle spectral properties to various cooperative phenomena in the
system of vortices\cite{franz98b,kwon99,balents}. 
The Doppler shift effect is not
peculiar to the vortex state. It also occurs in the Meissner
phase\cite{sauls} and is generally present whenever a quasiparticle
experiences a locally uniform drift in the superfluid velocity.  
{\em Second}, there is also a purely quantum effect 
which {\em is} intimately tied to
the vortex state: as a quasiparticle circles around a vortex while
maintaining its quantum coherence, the accumulated phase through a Doppler
shift is $\pm\pi$. This implies that there must be an {\em additional}
compensating $\pm\pi$ contribution to the phase on top of the one due to the 
Doppler shift\cite{franz1}.  The required $\pm\pi$ contribution is
supplied by a ``statistical interaction" or a
``Berry phase" effect and can be built in
at the Hamiltonian level as a half-flux Aharonov-Bohm scattering of
quasiparticles by vortices\cite{franz1}. 
This interplay between the classical (Doppler shift)
and purely quantum effect (``Berry phase") is what makes the problem of
quasiparticle-vortex interaction particularly fascinating. 

Let us briefly review what is already known about the subject.  The
initial theoretical investigations of gapped and gapless superconductors
in the vortex state were directed along rather separate lines.  The low
energy quasiparticle spectrum of an $s$-wave superconductor in the mixed
state was originally studied by Caroli, de~Gennes and Matricon (CdGM)
\cite{Caroli64} within the framework of the Bogoliubov-de~Gennes
equations\cite{deGennes89}. 
Their solution yields well known bound states in the vortex
cores. These states are {\em localized} in the core and have an
exponential envelope the scale of which is set by the BCS coherence
length. The low energy end of the spectrum is given by
$\epsilon_{\mu}\sim\mu (\Delta_0^2/E_F)$, where $\mu = 1/2,3/2,\dots$,
$\Delta_0$ is the overall BCS gap and $E_F$ is the Fermi energy. This
solution can be relatively straightforwardly generalized to a fully {\em
gapped}, {\em chiral} $p$-wave superconductor. In this case the low energy
quasiparticle spectrum also displays bound vortex core states, whose
energy quantization is, however, modified relative to its $s$-wave
counterpart, precisely because of the chiral character of a $p_{x \pm
iy}$-wave superconductor and the ensuing shift in the angular momentum.
For example, the low energy spectrum of quasiparticles in the singly
quantized vortex of the $p_{x \pm iy}$-wave superconductor, possesses a
state at exactly zero energy\cite{Volovikpw99,Matsumoto99}. 

By comparison, the spectrum of a {\em gapless} $d$-wave superconductor in
the mixed phase has become the subject of an active debate only relatively
recently, fueled by the interest in HTS.  Naturally, the first question
that arises is what is the analog of the CdGM
solution for a single vortex.  It is important to realize here that the
situation in a $d_{x^2-y^2}$ superconductor is {\em qualitatively} {\em
different} from the classic $s$-wave case\cite{franz0}: when the pairing
state has a finite angular momentum and is not a global eigenstate of the
angular momentum $L_z$ (a $d_{x^2-y^2}$ superconductor is an equal
admixture of $L_z=\pm 1$ states), the problem of fermionic excitations in
the core {\em cannot} be reduced to a collection of decoupled 1D
dimensional eigenvalue equations for each angular momentum channel, the
key feature of the CdGM solution. Instead, all
channels remain coupled and one must solve a {\em full} 2D problem. The
fully self-consistent numerical solution of the BdG
equations\cite{franz0,resende98} reveals the most important physical
consequence of this qualitatively new situation: the vortex core
quasiparticle states in a pure $d_{x^2-y^2}$ superconductor are {\em
delocalized} with wave functions extended along the nodal directions.  The
low lying states have a continuous spectrum and, in a broad range of
parameters, do not seem to exhibit strong resonant behavior.  Obviously,
this is in sharp contrast with a discrete spectrum and true bound
quasiparticle states of the CdGM $s$-wave solution. We
expect the above qualitative results to hold for all unconventional
superconductors and within confines of the simple BdG equations, 
as long as there are nodes in the gap. 

A particularly important issue in this context is the nature of the
quasiparticle excitations at very low fields, in the presence of a vortex
lattice. This is a novel challenge since the spectrum starts as gapless at
zero field and at issue is the interaction of these low lying
quasiparticles with the vortex lattice. This problem has been addressed
via numerical solution of the tight binding model \cite{wang1}, a
numerical diagonalization of the continuum model \cite{kita1} and a
semiclassical analysis \cite{Volovikdw93}.  Gorkov, Schrieffer
\cite{gorkov} and, in a somewhat different context, Anderson
\cite{anderson}, predicted that the quasiparticle spectrum is described by
a Dirac-like Landau quantization of energy levels
\begin{equation}
E_n =\pm \hbar \omega_H \sqrt{n},\;\; \; n=0,1,... 
\label{anderson}
\end{equation}
where $\omega_H= \sqrt{2 \omega_c \Delta_0/ \hbar}$,
$\omega_c=eB/mc$ is the cyclotron frequency and $\Delta_0$ is the
maximum superconducting gap. The Dirac-like spectrum of Landau levels
arises from the linear dispersion of nodal quasiparticles at zero field.
This argumentation neglects the effect of spatially varying supercurrents
in the vortex array which were shown to strongly mix individual Landau
levels \cite{Melnikov}. 

Recently, Franz and Te\v{s}anovi\'c (FT) \cite{franz1} pointed out that the
low energy quasiparticle states of a $d_{x^2-y^2}$-wave superconductor in
a vortex state are most naturally described by strongly dispersive Bloch
waves. This conclusion was based on the particular choice of a singular
gauge transformation, which allows for the treatment of the uniform
external magnetic field and the effects produced by chiral supercurrents
on equal footing. The starting point was the Bogoliubov-de~Gennes (BdG) equation
linearized around a Dirac node.  By employing the singular gauge
transformation  FT mapped the original problem onto
that of a Dirac Hamiltonian in periodic vector and scalar potentials,
comprised of an array of an effective magnetic Aharonov-Bohm half-fluxes,
and with a vanishing overall magnetic flux per unit cell. The FT gauge
transformation allows use of standard band structure and other zero-field
techniques to study the quasiparticle dynamics in the presence of vortex
arrays, ordered or disordered. Its utility was illustrated in 
Ref.\ \cite{franz1} through
computation of the quasiparticle spectra of a square vortex lattice. 
A remarkable feature of these spectra is the persistence of the massless
Dirac node at finite fields and the appearance of the ``lines of
nodes" in the gap at large values of the anisotropy ratio
$\alpha_D=v_F/v_{\Delta} $, starting at $\alpha_D \simeq 15$.
Furthermore, the FT transformation directly reveals that a quasiparticle moving
coherently through a vortex array experiences not only a Doppler shift
caused by circulating supercurrents but also an {\em additional}, ``Berry
phase'' effect: the latter is a purely quantum mechanical phenomenon and is
absent from a typical semiclassical approach. Interestingly, the
cyclotron motion in Dirac cones is {\em entirely} caused by such
``Berry phase'' effect, which takes the form of a  half-flux Aharonov-Bohm
scattering of quasiparticles by vortices, and does {\em not} explicitly
involve the external magnetic field. It is for this reason that the
Dirac-like Landau level quantization is absent from the exact quasiparticle
spectrum.

Further progress was achieved by Marinelli, Halperin and Simon
\cite{marinelli} who presented a detailed perturbative analysis of the
linearized Hamiltonian of Ref.\ \cite{franz1}. 
They showed that the presence of the particle-hole
symmetry is of key importance in determining the nature of the spectrum of
low energy excitations. If the vortices are arranged in a Bravais lattice,
they showed that, to all orders in perturbation theory, the Dirac node is
preserved at finite fields, i.e the quasiparticle spectrum remains gapless
at the $\Gamma$ point. This result masks intense mixing of individual
basis vectors (in the case of Ref. \cite{marinelli} these are Dirac plane
waves), including strong mixing of states far removed in energy. The
continuing presence of the massless Dirac node at the
$\Gamma$ point after the application of the external
field is thus not due to the lack of scattering
which is actually  remarkably strong. Rather, it is dictated by symmetry: 
Marinelli {\em et al.} demonstrated that the crucial agent responsible for
the presence of the Dirac node is the particle-hole symmetry, present at
every point in the Brillouin zone. The fact that it is the particle-hole symmetry
rather than the lack of scattering that protects the Dirac node 
is clearly revealed in the related
problem of a Schr\"odinger electron in the presence of a single
Aharonov-Bohm half-flux, where the density of states acquires a $\delta$
function depletion at $\bk=0$ \cite{moroz}, thus shifting part of the
spectral weight to infinity due to remarkably strong scattering.
The authors of Ref. \cite{marinelli} also corrected Ref. \cite{franz1} by showing
that the ``lines of nodes" must actually be the ``lines of
near nodes" since true zeroes of the energy away from Dirac node
are prohibited on symmetry grounds. Still, these ``lines"
will act as true nodes in all realistic circumstances, 
due to extraordinarily small excitation energies.

Marinelli {\em et al.} also showed that, if the particle-hole symmetry is
broken, for example by introducing a non-Bravais vortex lattice with
broken inversion symmetry and four vortices in the unit cell, then true lines
of nodes can develop for values of anisotropy ratio
starting already at $\alpha_D \simeq 5$. They
concluded, that the density of states is finite at zero energy and the
semiclassical results of Kopnin and Volovik\cite{Kopnin96} might apply down to 
zero energy. For a non-Bravais lattice with two vortices per unit cell they found that the
quasiparticle spectrum can become gapped. 

Very recently Ye \cite{ye1} 
discussed transport properties of the quasiparticles described by the Dirac
Hamiltonian of Ref.\ \cite{franz1} and pointed out some intriguing effects
that may take place in random vortex arrays. Also, Altland,
Simons and Zirnbauer investigated general properties
of disordered Dirac operators, including vortex disorder\cite{zirn}.

In this paper we extend the original analysis which was based solely on the
{\em continuum} description by introducing a tight binding
``regularization'' of the full lattice BdG Hamiltonian,  
to which we then apply the FT gauge transformation. Our motivation
is twofold: First, we have found by explicit numerical computations that
different choices of singular gauge transformation result in spectra
which, while rather similar, are not the same.
Within our numerical accuracy we could not tell whether the 
spectra have a very slow convergence to the same final result or
whether they actually converge to a different answer. This will
be discussed again shortly. This problem appears to be a conspiracy between the
strong Aharonov-Bohm scattering from 
magnetic half-fluxes which tends to push some states of the
unperturbed Hamiltonian to very high energies
and the unbounded nature of the Dirac spectrum.
It is an interesting issue for future study how to devise the cutoff
in the reciprocal lattice sums of the linearized problem
which is tailor-made for a particular
choice of the singular gauge transformation.
In this paper, we circumvent this problem entirely by regularizing
the original Hamiltonian on a square lattice. 
The tight binding formulation regularizes the strong
mixing of the basis vectors 
through the introduction of an upper and a lower bound on the spectrum,
thus prohibiting the shift of the spectral weight to infinity
\cite{moroz}. This immediately solves our problem: different choices of
singular gauge transformation now rapidly converge to identical spectra,
as they should. The low energy part of the spectrum compares best with
the original FT transformation\cite{franz1} of the linearized Hamiltonian,
which might have been expected based on its having the smoothest relative
phase between particles and holes.

Second, the lattice formulation allows us to 
study what, if any,  role is played by {\em internodal} scattering 
which is simply not a part of
the linearized description. We find that under {\em special} circumstances,
when there is a high degree of commensurability between the ionic and
vortex lattices, the interference between the nodes can lead to scattering 
which is surprisingly strong ($\sim \sqrt{B}$) and might
be observable in HTS. Such scattering is responsible for
opening a gap at the Fermi surface even in the case a Bravais vortex lattice.
In a {\em typical} situation, however, when the
two lattices have a low degree of commensurability or are of different
symmetry and particularly when weak thermal or quenched
disorder is included, the internodal scattering effectively disappears.
We diagonalize the tight binding Hamiltonian numerically 
for various order parameter symmetries
and both square and triangular vortex lattices.  Our treatment provides an 
access to the entire quasiparticle energy spectrum together with displaying the
utility of the FT transformation in analyzing gapped superconductors (e.g.
$s$- or $p_{x+iy}$- wave), which are {\it a priori} inaccessible through
the linearization. 
We are therefore able to present a unified treatment of
a general, both conventional and unconventional,
strongly type-II superconducting pairing in the vortex state. 

%
%


\section{BdG Hamiltonian and the Singular Gauge Transformation: Low
Energy Physics of Quasiparticles and Vortices}

Because of the nonlocality inherent in the superconductors with 
higher angular momentum pairing, their Hamiltonians are most naturally
formulated on a discrete real space lattice representing the underlying 
crystalline lattice of the compound in question. Quite generically, the 
simplest lattice Hamiltonian which allows pairing to occur in $s$-, $p$- and
$d$-wave channels is the tight binding model with the on-site or nearest 
neighbor attraction between electrons. Conventional mean field 
Hartree-Fock-Bogoliubov
decoupling of the interaction term then leads to the BCS-type lattice
Hamiltonian of the form 
\begin{equation}
 {\cal H}=\left( \begin{array}{cc}
\hat{h} & \hat{\Delta} \\
\hat{\Delta}^{\ast} & -\hat{h}^{\ast}
\end{array} \right)
\label{IIi}
\end{equation}
where
\begin{equation} 
\hat{h} = -t \sum_{{\bf \delta}}e^{-i \frac{e}{\hbar c}\int_{{\bf r}}^{{\bf r} 
+ {\bf \delta}}{\bf A}({\bf r})\cdot d{\bf l}}\;\hat{s}_{{\bf \delta}}-
\epsilon_F
\end{equation}
and 
\begin{equation}
\hat{\Delta}= \Delta_{0} \sum_{{\bf \delta}}e^{i \phi({\bf r})/2}\; 
\hat{\eta}_{{\bf \delta}} \;e^{i \phi({\bf r})/2}
\label{IIiii}
\end{equation}
The sums are over nearest neighbors and on the square lattice ${\bf
\delta}= \pm \hat{x}, \pm \hat{y}$; ${\bf A}({\bf r})$ is the vector
potential associated with the external magnetic field $B$, $\epsilon_F$ is
Fermi energy, and
$\hat{s}_{{\bf \delta}}$ is an operator which is defined by its action on
a general function $u({\bf r})$ so that $\hat{s}_{{\bf \delta}} \,u({\bf
r}) =u({\bf r}+{\bf \delta})$. The operator $\hat{\eta}_{{\bf \delta}}$
depends on the type of pairing as discussed later. 

Quasiparticle wavefunction is a rank two spinor in the Nambu space, 
${\psi}^T({\bf r})=[u({\bf r}),v({\bf r})]$, and  obeys 
the BdG equation 
\begin{equation}
{\cal H}\psi = \epsilon\psi.
\label{bdg}
\end{equation}

Besides relying on conventional mean-field BCS decoupling, Hamiltonian
(\ref{IIi}) contains two additional approximations. First, we have assumed
that the order parameter magnitude is constant and equal to $\Delta_0$
everywhere in space. This is essentially the London limit\cite{deGennes89}
which is expected to
be valid in the regime of low fields, $B\ll H_{c2}$, when vortex cores 
comprise negligible fraction of the sample. Second, we approximated the phase
of the order parameter $\phi_{\bf \delta}(\br)$, 
which is a nonlocal field associated with a {\em bond}
between two neighbor sites, by the average of the phases associated with the
attached sites,
\begin{equation}
\phi_{\bf \delta}(\br)\to {1\over 2}[\phi(\br)+\phi(\br+{\bf \delta})].
\label{phs}
\end{equation}
This replacement is discussed in more detail in the Appendix A and 
we expect it to be very accurate far away from the vortex cores
where the phase varies slowly, but inadequate in the core. 
Hamiltonian (\ref{IIi}) is therefore useful when considering quasiparticle
properties in a dilute vortex lattice, which is the main focus of this work. 
To study properties of the core region one must explicitly treat the order 
parameter amplitude variation and nonlocality of its phase as done e.g.
in Refs.\ \cite{franz0,wang1}. Surprisingly, however, we shall see below
that even the present approximation yields results for the core region that 
are qualitatively correct.

\subsection{Continuum formulation} 

In many cases our main interest is directed at the long wavelength and low 
energy or low
temperature properties. It is precisely in this respect that the quasiparticle
excitations in an unconventional superconductor differ most
dramatically from its $s$-wave counterpart. Under these circumstances it is
desirable to consider
 a continuum version of the BdG Hamiltonian. For a $d$-wave 
superconductor such a continuum Hamiltonian was derived by Simon and 
Lee\cite{sl1}. It turns out, however, that as written in Ref.\ \cite{sl1}
this Hamiltonian is not gauge invariant\cite{remark1}. At fault is the
off-diagonal term representing the $d$-wave pairing operator, which does
not transform properly under the U(1) gauge group. 
In Appendix A we have derived the gauge invariant form of this 
pairing operator for a pure $d_{x^2-y^2}$
superconductor and have outlined how such a derivation can be carried out
for other unconventional pairing states. The continuum Hamiltonian
reads:
\begin{equation}
{\cal H} =\left( \begin{array}{cc}
\hat{\cal H}_e      & \hat{\Delta} \\
\hat{\Delta}^*  & -\hat{\cal H}_e^*
\end{array} \right),
\label{h1}
\end{equation}
with $\hat{\cal H}_e={1\over 2m}({\bf p}-{e\over c}{\bf A})^2-\epsilon_F$ and
$\hat{\bf p}=-i\hbar\nabla$ the momentum operator.
If we follow the convention and choose the coordinate axes in the direction 
of gap nodes the {\em gauge invariant} $d$-wave pairing operator has the 
form 
\begin{equation}
\hat\Delta=p_F^{-2}\{\hat p_x,\{\hat p_y,\Delta({\bf r})\}\}
+{i\over 4}p_F^{-2}\Delta({\bf r})(\hat p_x\hat p_y \phi),
\label{delta1}
\end{equation}
where $p_F$ is the Fermi momentum, $\phi$ is the phase of the superconducting
gap $\Delta({\bf r})$ and curly brackets represent
symmetrization, $\{a,b\}={1\over 2}(ab+ba)$.
The above pairing operator resembles the familiar Simon-Lee form except
for the last term which is necessary to restore the full gauge invariance.
We emphasize that expression (\ref{delta1}) is valid for uniform gap 
amplitude; otherwise additional terms which involve derivatives of the 
amplitude appear.

We now use this Hamiltonian as
the starting point in our discussion of low energy quasiparticles in the
presence of magnetic field. Operationally, the main difficulty encountered
when solving for eigenstates of (\ref{h1}) in the vortex state is the
nontrivial structure of the order parameter phase field $\phi(\br)$, 
which is constrained
by topology to wind by $2\pi$ around the center of each vortex.
Ideally, we would want to get rid of this phase to
make the problem look as close as possible to the reference solution
in which the phase can simply be set to zero. If $\phi ({\bf r})$
is a pure gauge, i.e. $\nabla\times\nabla\phi ({\bf r}) = 0$, this is
easily accomplished by performing a gauge transformation
\begin{equation}
{\cal H} \to U^{-1}{\cal H}U, \ \ \ 
U=\left( \begin{array}{cc}
e^{i\phi({\bf r})/2}      & 0 \\
0  & e^{-i\phi({\bf r})/2} 
\end{array} \right).
\label{u1}
\end{equation}
After this transformation the BdG Hamiltonian becomes
\[
\left( \begin{array}{cc}
{1\over 2m}(\hat{\bf p}+m{\bf v}_s)^2-\epsilon_F & {\Delta_0\over p_F^2}
\hat p_x\hat p_y \\
{\Delta_0\over p_F^2}\hat p_x\hat p_y  & 
-{1\over 2m}(\hat{\bf p}-m{\bf v}_s)^2+\epsilon_F
\end{array} \right),
\]
where 
\begin{equation}
{\bf v}_s({\bf r})={1\over m}({\hbar\over 2}\nabla\phi-{e\over c}{\bf A})
\end{equation}
is the conventional superfluid velocity.
We recognize the term containing $\nabla\phi$ as the Doppler shift
of quasiparticles in a locally uniform superflow\cite{Volovikdw93,sauls}.

However, if  $\phi ({\bf r})$ contains vortices the situation is far more
subtle: the vector field $\nabla\phi ({\bf r})$, while still locally
uniform, acquires a {\em global} curvature, i.e. 
\begin{equation}
\nabla\times\nabla\phi ({\bf r}) = 2\pi\hat{z}\sum_i
\delta ({\bf r} -{\bf r}_i)\neq 0,
\label{phase1}
\end{equation}
where $\{\br_i\}$ denotes vortex positions. Consequently, in the vortex state 
it is no longer possible to eliminate the superconducting phase by the
above transformation (\ref{u1}) and obtain a Hamiltonian describing 
quasiparticles coupled to the locally uniform superflow. Formally this can be
seen from the fact that  in the 
presence of vortices transformation (\ref{u1}) is not single valued. In
principle such multiple valuedness of the resulting Hamiltonian could be
handled by introducing compensating branch cuts in the quasiparticle
wavefunctions. In practice, however, it is far more desirable to avoid
any such complications in the first place.

We follow FT \cite{franz1} and perform a ``bipartite'' singular gauge 
transformation,
\begin{equation}
{\cal H} \to U^{-1}{\cal H}U, \ \ \ 
U=\left( \begin{array}{cc}
e^{i\phi_e({\bf r})}      & 0 \\
0  & e^{-i\phi_h({\bf r})} 
\end{array} \right),
\label{u2}
\end{equation}
where $\phi_e({\bf r})$ and $\phi_h({\bf r})$ are two functions 
satisfying
\begin{equation}
\phi_e({\bf r})+\phi_h({\bf r})=\phi({\bf r}).
\label{con1}
\end{equation}
This more general transformation also eliminates the phase of the order 
parameter from the pairing term of the Hamiltonian but 
$\phi_e({\bf r})$ and $\phi_h({\bf r})$ now can be chosen in a way that avoids
multiple valuedness and the associated complications. The 
way to accomplish this is to assign the singular part of the phase field 
generated by any given vortex to either $\phi_e({\bf r})$ or 
$\phi_h({\bf r})$, but not both as is done by symmetric transformation
(\ref{u1}). Physically, a vortex assigned to $\phi_e({\bf r})$ will be seen 
by electrons and be invisible to holes, while vortex assigned to 
$\phi_h({\bf r})$ will be seen holes and be invisible to electrons.
\begin{figure}[t]
\epsfxsize=8.5cm
\hfil\epsfbox{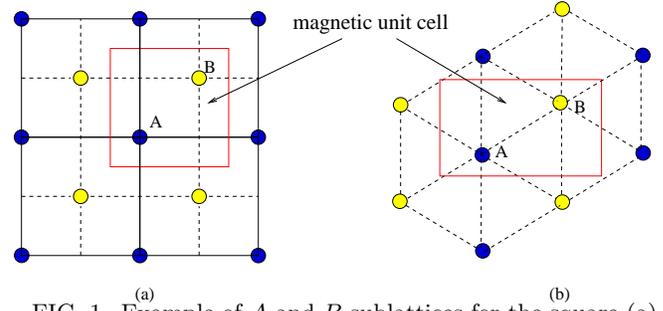}\hfill
\caption{Example of $A$ and $B$ sublattices for  the square (a) and 
triangular (b) vortex lattice.}
\label{transf}
\end{figure}

In practice we implement the above transformation by dividing vortices into
two groups $A$ and $B$, positioned at $\{\br_i^A\}$ and $\{\br_i^B\}$
respectively (see Fig.~\ref{transf}). We then define two phase fields 
$\phi_A({\bf r})$ and  $\phi_B({\bf r})$ such that
\begin{equation}
\nabla\times\nabla\phi_\mu({\bf r}) = 2\pi\hat{z}\sum_i
\delta ({\bf r} -{\bf r}_i^\mu), \ \ \ \mu=A,B,
\label{del}
\end{equation}
and identify $\phi_e=\phi_A$ and $\phi_h=\phi_B$. Comparison with Eq.
(\ref{phase1}) confirms that this choice of $\phi_e({\bf r})$ and 
$\phi_h({\bf r})$ satisfies the condition (\ref{con1}), possibly up to some
unimportant nonsingular phase which can be always transformed away by 
the conventional gauge transformation (\ref{u1}). 

The transformed Hamiltonian becomes\cite{remark2}
\[
\left( \begin{array}{cc}
{1\over 2m}(\hat{\bf p}+m{\bf v}_s^A)^2-\epsilon_F & \hat D \\
\hat D  & 
-{1\over 2m}(\hat{\bf p}-m{\bf v}_s^B)^2+\epsilon_F
\end{array} \right),
\]
with $\hat D={\Delta_0\over 2p_F^2}[\hat p_x+{m\over 2}(v_{sx}^A-v_{sx}^B)]
[\hat p_y+{m\over 2}(v_{sy}^A-v_{sy}^B)] + (x\leftrightarrow y)$ and 
\begin{equation}
{\bf v}_s^\mu={1\over m}(\hbar\nabla\phi_\mu-{e\over c}{\bf A}), \ \ \mu=A,B.
\label{vsab}
\end{equation}
~From the perspective of quasiparticles ${\bf v}_s^A$ and ${\bf v}_s^B$ 
can be thought of as
{\em effective} vector potentials acting on electrons and holes respectively. 
Corresponding effective magnetic field seen by the quasiparticles is
${\bf B}_{\rm eff}^\mu=-{mc\over e}(\nabla\times{\bf v}_s^\mu)$, and can be 
expressed using Eqs.\ (\ref{del}) and (\ref{vsab}) as  
\begin{equation}
{\bf B}_{\rm eff}^\mu=\bB-\phi_0\hat{z}\sum_i
\delta ({\bf r} -{\bf r}_i^\mu), \ \ \ \mu=A,B,
\label{beff}
\end{equation}
where $\bB=\nabla\times\bA$ is the physical magnetic field and 
$\phi_0=hc/e$ is the flux quantum. We observe that quasi-electrons and 
quasi-holes propagate in the effective field which consists of 
(almost) uniform 
physical magnetic field $\bB$ and an array of opposing delta function 
``spikes'' of unit fluxes associated with vortex singularities. The latter
are different for electrons and holes.
As discussed in \cite{franz1} it is desirable to choose $A$ and 
$B$ vortices in such a way that the effective magnetic field 
vanishes on average, i.e. $\langle{\bf B}_{\rm eff}^\mu\rangle=0$.
This translates to a simple requirement that we have precisely one flux spike
(of $A$ and $B$ type) per flux quantum of the physical magnetic field. In that
case flux quantization guarantees that the right hand side of Eq.\ 
(\ref{beff}) vanishes when averaged over a vortex lattice unit cell containing
two physical vortices. It also implies that there must be equal numbers of 
$A$ and $B$ vortices in the system. 

The essential advantage of the choice with vanishing 
$\langle{\bf B}_{\rm eff}^\mu\rangle$ is
that  ${\bf v}_s^A$ and ${\bf v}_s^B$ can be chosen periodic in space
with periodicity of the magnetic unit cell containing one electronic
flux quantum $hc/e$. Notice that vector potential of a field that does not
vanish on average can never be periodic in space. Condition 
$\langle{\bf B}_{\rm eff}^\mu\rangle=0$
is therefore crucial in this respect. 

The singular gauge transformation (\ref{u2}) maps the original Hamiltonian
of fermionic quasiparticles in finite magnetic field onto a new Hamiltonian
which is formally in zero average field and has no singular phase winding
in the off-diagonal components. The main advantage of the FT transformation
is that it eliminates the need to introduce branch cuts into fermionic
wavefunctions: these wavefunctions remain single valued while the physical
effect of the branch cuts is now promoted to the level of the Hamiltonian,
where it is represented by a statistical gauge potential describing
a half-flux Aharonov-Bohm scattering.
This situation bears some similarity to 
the fractional quantum Hall effect (FQHE). Here, the composite fermion
\cite{jain1,halperin1} is created by attaching a flux tube to the electron. 
The details, however, are quite different.
In the present case it is the superconducting condensate that creates the 
fictitious ``flux spikes'' which then on average exactly neutralize the
physical applied magnetic field. Unlike in FQHE, the fluxes are stationary
and we are generally in the limit where there is a large number of electrons 
per flux.

To facilitate further insights into the physics of the 
low-energy quasiparticles
we now linearize the transformed Hamiltonian in the vicinity of one of the
four nodes of the gap function on the Fermi surface. Following Simon
and Lee\cite{sl1} we obtain ${\cal H}_N\simeq {\cal H}_0+
{\cal H}'$, where
\begin{equation}
{\cal H}_0 =\left( \begin{array}{cc}
v_F\hat p_x & v_\Delta\hat p_y \\
v_\Delta\hat p_y & -v_F\hat p_x
\end{array} \right)
\label{h0}
\end{equation}
is the free Dirac Hamiltonian and 
\begin{equation}
{\cal H}' =m\left( \begin{array}{cc}
v_Fv_{sx}^A & {1\over 2}v_\Delta(v_{sy}^A-v_{sy}^B) \\
{1\over 2}v_\Delta(v_{sy}^A-v_{sy}^B) & v_Fv_{sx}^B
\end{array} \right).
\label{h'}
\end{equation}
Here $v_F$ is the Fermi velocity and 
$v_\Delta=\Delta_0/p_F$ denotes the slope of the gap at the node.

${\cal H}_N$ can be viewed as a relativistic Hamiltonian for a 2+1 massless
``Dirac'' fermion and can be rewritten accordingly as 
\begin{equation}
{\cal H}_N=v_F(\hat{p}_x +a_x)\tau_3 + v_\Delta(\hat{p}_y +a_y)\tau_1
+mv_F v_{sx}~,
\label{h''}
\end{equation}
where $\tau_i$ are Pauli matrices, ${\bf v}_s={1\over 2}({\bf v}_s^A
+{\bf v}_s^B)={1\over m}({\hbar\over 2}\nabla\phi-{e\over c}{\bf A})$ is the
conventional superfluid velocity and ${\bf a}={m\over 2}({\bf v}_s^A
-{\bf v}_s^B)={\hbar\over 2}(\nabla\phi_A-\nabla\phi_B)$ is the internal
gauge field. We observe that ${\bf v}_s$ couples to the Dirac fermions
as a {\em scalar} potential while ${\bf a}$ couples as a {\em vector} 
potential. The Dirac ``magnetic field'' ${\bf b}=\nabla\times{\bf a}$ produced 
by this vector potential is highly unusual: it consists of delta function
spikes located at the vortex centers and it vanishes on average when the
numbers of $A$ and $B$ vortices are equal. Each spike carries precisely 
one half of the conventional electronic flux quantum $\phi_0$ and therefore,
although comprising a set of measure zero in the real space, the flux spikes
lead to {\em maximal} Aharonov-Bohm scattering and have strong effect on
the quasiparticle spectra.  In particular, note that the cyclotron motion in a Dirac
cone arises {\em entirely} through ${\bf b}=\nabla\times{\bf a}$ and
does {\em not} include explicitly the actual magnetic field
${\bf B}=\nabla\times{\bf A}$. Such half-flux scattering is time-reversal
invariant and cannot lead to Dirac-like (or any!) Landau level
quantization.


\subsection{Internal gauge symmetry}

Spectral properties of the continuum linearized Hamiltonian (\ref{h0}-\ref{h'})
have been analyzed in great detail\cite{franz1,marinelli} and initial 
investigation of its transport properties has been presented\cite{ye1}.
Here we wish to point out a peculiar property of the linearized Hamiltonian
as regards the choice of $A$ and $B$ subsets of vortices, which seems
to have been overlooked thus far. 

Logic dictates that all measurable quantities must be independent of our 
choice of $A$ and $B$. This is because there should be no physical
distinction between $A$ and $B$ vortices, the assignment being completely
arbitrary. The freedom of assignment of 
vortices into $A$ and $B$ subsets represents an internal gauge symmetry of the 
problem closely related to the fact that electrons condense in pairs and 
therefore vortices carry {\em half} of the electronic flux quantum $hc/e$.
\begin{figure}[t]
\epsfxsize=8.5cm
\hfil\epsfbox{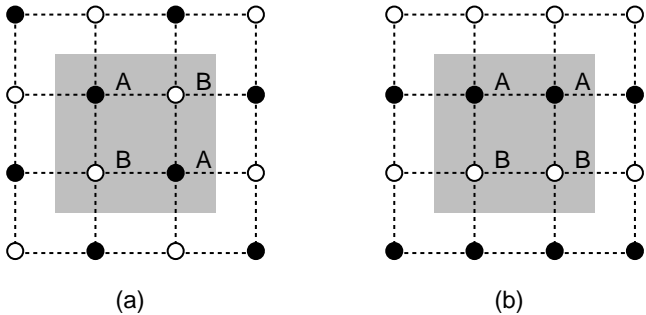}\hfill
\caption{Two sublattices, ``$ABAB$'' and ``$AABB$'' used to investigate the
internal gauge symmetry of the FT transformation. The shaded region 
marks the unit cell used in the numerical diagonalization.
}
\label{subla}
\end{figure}

\begin{figure}[t]
\epsfxsize=8.5cm
\hfil\epsfbox{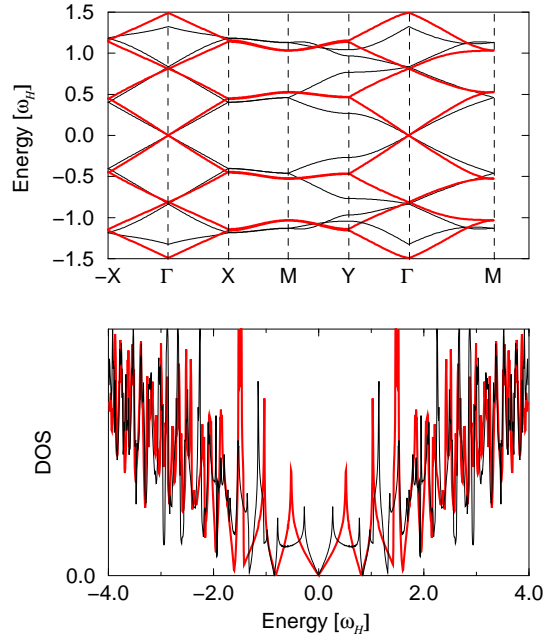}\hfill
\caption{The band structure (top) and DOS (bottom) of the linearized 
Hamiltonian with two choices of sublattices: $ABAB$ (thick line) and
$AABB$ (thin line). 
}
\label{abcomp}
\end{figure}

To explicitly test this internal gauge symmetry we have diagonalized the 
linearized 
Hamiltonian (\ref{h0}-\ref{h'}) using the Bloch wave technique described 
in Ref.\ \cite{franz1} for two distinct choices of $A$-$B$ sublattices as
illustrated in Fig. \ref{subla}. We used a unit cell containing 4 vortices 
in order to be able to compare the band structures for the two cases 
directly on the same Brillouin zone. The corresponding band structures 
for the isotropic case ($\alpha_D=v_F/v_\Delta=1$) are
displayed in Fig.\ \ref{abcomp}. We observe that although qualitatively 
similar, their detailed features are {\em different} and so are the 
associated densities of states. Similar situation occurs for other values
of Dirac cone anisotropy $\alpha_D$ and other symmetries of the vortex 
lattice, although the case shown in Fig.\ \ref{abcomp} is an extreme 
example of the differences. This is a surprising and unexpected 
result whose ramifications we do not fully understand at the present time. 

We expended considerable effort to verify that the difference between the 
two band structures is not a trivial artifact of our diagonalization 
procedures. Rather, it appears to be associated with the pathological 
$\sim r^{-1/2}$
behavior of the Dirac wavefunctions in the vicinity of a vortex center, which 
is presumably difficult to mimic using finite number of Bloch waves that are
used as basis states in our numerical diagonalization. We also note that 
the Aharonov-Bohm scattering induced by the half-flux
spikes is extraordinarily strong. As shown by Moroz\cite{moroz}, in the case
of ordinary Schr\"odinger electron it causes a transfer of spectral weight 
from zero energy up to infinite energy. 

The problem is clearly inherent only
to the {\em linearized} BdG Hamiltonian. In the following Section we show 
that no such problem arises in the lattice version of the BdG Hamiltonian. 
This is
presumably because the spectrum is bounded (by the tight binding bandwidth)
in this case and therefore all states are accounted for in the numerical
diagonalization. Also, the lattice spacing acts as a natural short distance
cutoff which regularizes the behavior of the wavefunctions at the core. 

We have also solved the linearized problem by directly discretizing the 
Hamiltonian (\ref{h0}-\ref{h'}) on a square grid in the real space, a 
technique similar to that described in Ref.\ \cite{marinelli}. The problem 
persists in this case. We conclude that the problem appears to be a caused by
a conspiracy between the strong  Aharonov-Bohm scattering and the
unbounded nature of the Dirac spectrum of Hamiltonian (\ref{h0}-\ref{h'}). 
While we believe that there 
exists a regularization scheme which would resolve the problem within the
linearized formulation, our attempts to construct such a scheme were
unsuccessful so far. We leave it as an interesting problem for further
investigation.


\subsection{Lattice formulation}

It is straightforward to apply the FT singular gauge transformation (\ref{u2})
to the lattice BdG Hamiltonian (\ref{IIi}). One obtains Hamiltonian 
${\cal H}_N$ of the form 
\begin{equation}
\left( \begin{array}{cc}
-t \sum_{{\bf \delta}}
e^{i{\cal V}_{\bf\delta}^A(\br)}
\hat{s}_{{\bf \delta}}-\epsilon_F \; &
 \Delta_{0} \sum_{{\bf \delta}}e^{-\frac{i}{2}\delta\phi}\
 \hat{\eta}_{{\bf \delta}} \;e^{\frac{i}{2}\delta\phi} \\
 \Delta_{0} \sum_{{\bf \delta}}e^{-\frac{i}{2}\delta\phi}
 {{\hat{\eta}}^{\ast}}_{{\bf \delta}} \;e^{\frac{i}{2}\delta\phi} &
t \sum_{{\bf \delta}}e^{-i{\cal V}_{\bf\delta}^B(\br)}
\hat{s}_{{\bf \delta}}+\epsilon_F\;
\end{array} \right)
\end{equation}
where
\begin{equation}
{\cal V}_{\bf\delta}^\mu(\br)=\int_{{\bf r}}^{{\bf r} + 
{\bf \delta}}(\nabla \phi_\mu - \frac{e}{\hbar c} {\bf A})\cdot d{\bf l}~,
\ \ \ \ \mu=A,B,
\label{calv}
\end{equation}
and
\begin{equation}
\delta\phi=\phi_A-\phi_B.
\end{equation}
We notice that the integrand of Eq.\ (\ref{calv}) is proportional to the 
superfluid
velocities $\bv_s^\mu$ defined by Eq.\ (\ref{vsab}) in connection with the
continuum Hamiltonian. Appendix B describes an efficient method for 
calculation of these quantities in the vortex lattice. 

The main benefit of re-framing the original problem in this way
is the explicit gauge invariance.
In the case of a periodic arrangement of vortices 
the Hamiltonian is periodic with periodicity of a magnetic unit cell 
containing a pair of $A$ and $B$ vortices. In what follows we consider
square and triangular vortex lattices with two physical vortices per unit
cell as illustrated in Fig.\ \ref{transf}. The vortex center is always
placed at the center of the plaquette of the underlying tight binding
lattice.

Following FT we use the familiar Bloch 
states as the natural basis for finding the eigenvalues of  ${\cal H}_N$
specified above. In particular we seek the eigensolution of the BdG
equation ${\cal H}_N\psi=\epsilon\psi$ in the Bloch form
\begin{equation}
\psi_{n\bk}(\br) = e^{i\bk\cdot\br}\Phi_{n\bk}(\br)=
e^{i\bk\cdot\br}\left( \begin{array}{c} U_{n\bk}(\br) \\
V_{n\bk}(\br) \end{array} \right),
\label{bloch}
\end{equation}
where $(U_{n\bk},V_{n\bk})$ are periodic on the corresponding unit cell, 
$n$ is a band 
index and $\bk$ is a wave vector from the first Brillouin zone. Bloch
wavefunction  $\Phi_{n\bk}(\br)$ satisfies the ``off-diagonal'' Bloch 
equation 
$${\cal H}_{\bk}\Phi_{n\bk}=\epsilon_{n\bk}\Phi_{n\bk},$$
with the
Hamiltonian of the form ${\cal H}_{\bk}=e^{-i\bk\cdot\br}{\cal H}_N
e^{i\bk\cdot\br}$. In the following Section we describe specific forms of 
such Hamiltonians for the $s$-, $p$- and $d$-wave symmetries and discuss
their solutions.

\section{Numerical results}

\subsection{$s$-wave pairing}

In the case of $s$-wave pairing the operator $\hat{\eta}_{{\bf \delta}}$ 
takes the form $\hat{\eta}_{{\bf \delta}}=\frac{1}{4}$, and
the Hamiltonian simplifies considerably. In particular,
 the off-diagonal terms become simply $\Delta_0$, and ${\cal H}_N$ reads
\begin{eqnarray}
\left( \begin{array}{cc}
-t \sum_{{\bf \delta}}e^{i {\cal V}_{\bf\delta}^A(\br)}
\hat{s}_{{\bf \delta}}- \epsilon_F &
 \Delta_{0} \\
 \Delta_{0} & t \sum_{{\bf \delta}}e^{-i {\cal V}_{\bf\delta}^B(\br)}
\hat{s}_{{\bf \delta}}+ \epsilon_F
\end{array} \right).
\end{eqnarray}
It is interesting to note that
in the limit of high quasiparticle energy, $\epsilon \gg {\Delta}_0$,
the off-diagonal terms become irrelevant, and the equations for 
the electron and hole part of the Nambu wavefunction decouple.
We recover a Hamiltonian describing holes and electrons 
in a uniform magnetic field pierced by a lattice of 
counteracting full Aharonov-Bohm magnetic flux tubes with unit flux
quanta ${h c}/{e}$ concentrated at the set of point cores. 
The solution is just the familiar Schr\"odinger Landau levels (not to be
confused with Eq.~\ref{anderson}) because the full electronic 
flux has no effect on the
particle energy spectrum \cite{nielsen}. This result is expected from the 
outset since at high 
energies the quasiparticles behave as normal electrons or holes, 
which know little about the condensate. These high energy quasiparticles 
experience effectively a uniform magnetic field and move along 
cyclotron orbits. Similar argument holds for any pairing symmetry 
and we expect Landau level quantization of the quasiparticle spectrum at 
energies much larger than $\Delta_0$.
\begin{figure}[t]
\epsfxsize=4.5cm
\hfil\epsfbox{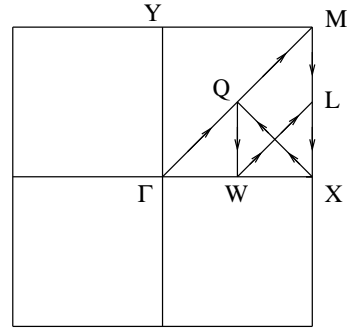}\hfill
\caption{Magnetic Brillouin zone for the square vortex lattice 
with the corresponding notations used in the discussion of the quasiparticle 
band structure.}
\label{brilzone}
\end{figure}

\begin{figure}
\epsfxsize=8.5cm
\hfill\epsfbox{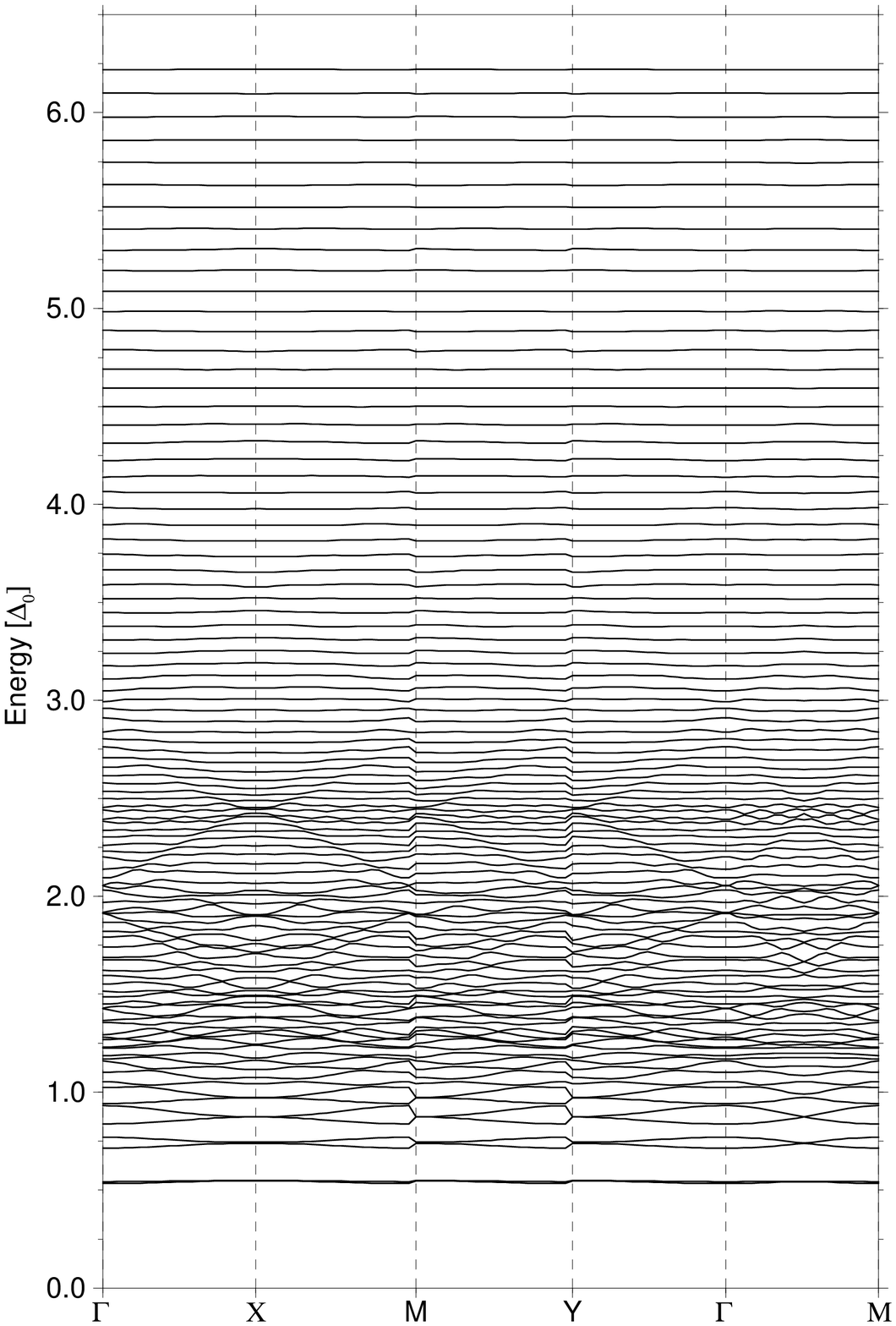}\hfill
\vspace{0.5cm}
\epsfxsize=8.5cm
\hfill\epsfbox{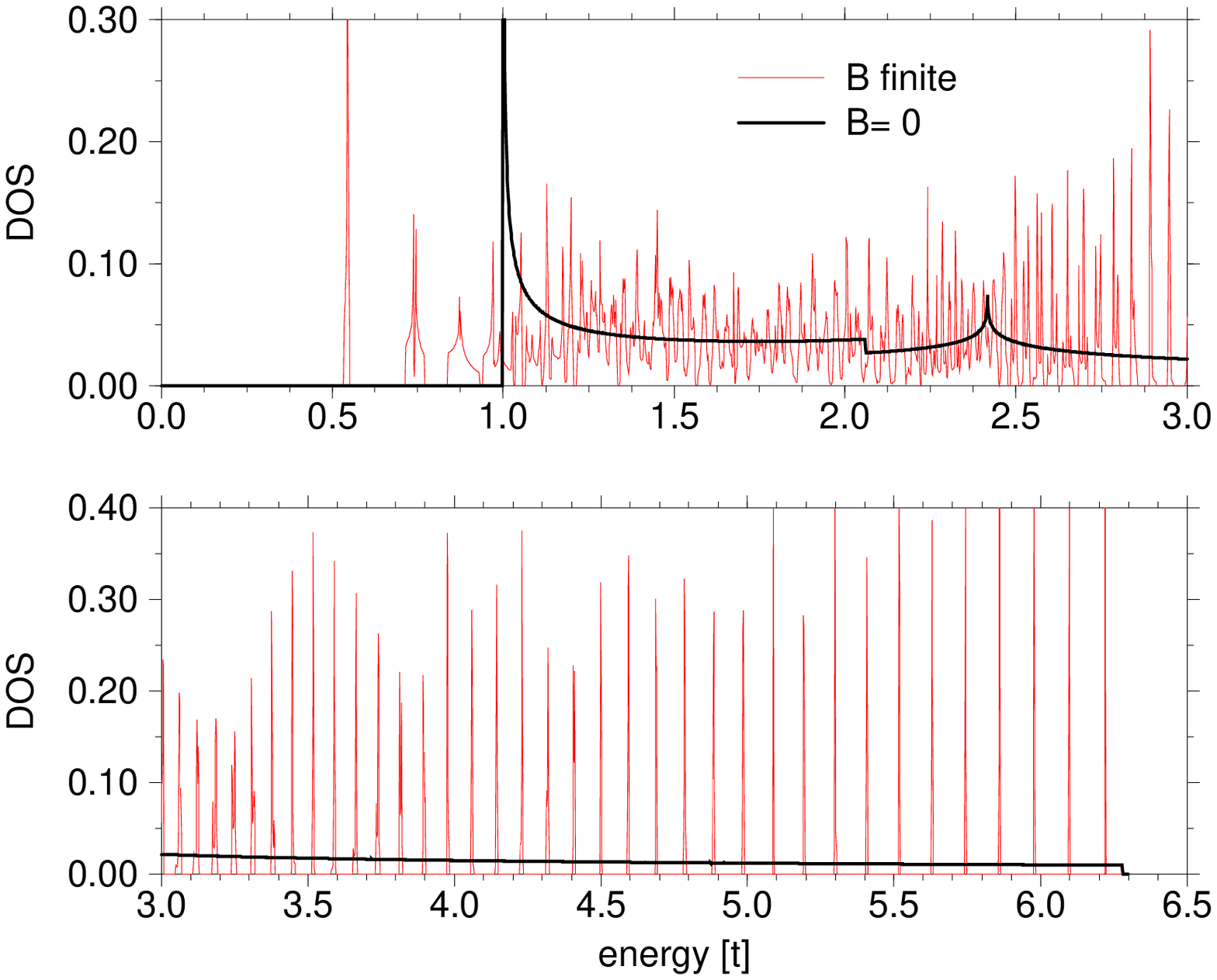}\hfill
\caption{Top: Quasiparticle band spectrum for an $s$-wave superconductor in the
presence of the external magnetic field B=1/100$\delta^2$, and $\Delta_0=t$, 
$\epsilon_F=-2.2t$. Bottom: corresponding DOS. Note the bound Caroli-Matricon 
bands at energies below the gap and the Landau levels at energies 
$\epsilon \gg \Delta_0$.}
\label{swaveband}
\end{figure}

We have 
numerically diagonalized the above Hamiltonian making use of the standard 
LAPACK diagonalization routine. We considered a tight-binding lattice of 
$10\times 10$ sites, 
which turns out to be sufficiently large to analyze the 
CdGM regime. The corresponding 
magnetic field $B=1/(100{\delta}^2)$ in units of unit flux $hc/e$ 
per unit area, the superconducting gap ${\Delta}_0=t$ and the chemical 
potential $\epsilon_F =-2.2t$, assuring an approximately cylindrical Fermi surface. 
The resulting spectrum for the Brillouin zone displayed in Fig.\ \ref{brilzone}
and density of states for the square vortex lattice
are shown in the Fig.~\ref{swaveband}. The $B=0$ spectrum has the usual
BCS form with a full gap $\Delta_0$. The additional features at $2.1\Delta_0$
and $2.4\Delta_0$ are remnants of the band edge and the van Hove singularity
respectively present in the normal state spectrum. 

The magnetic field induces low-energy states within the gap, 
which become localized in the vortex cores. 
These are CdGM states \cite{Caroli64} dispersed into bands. 
At low energies, the bands are very narrow signaling strong concentration 
of the 
wavefunctions at the vortex cores and insignificant overlaps among the 
states at neighboring vortices. This fact justifies the chosen parameters. 
At energies less, but comparable to $\Delta_0$, 
the bands are broadened due to increasing overlap among the wavefunctions. 
In is interesting to note that CdGM bound states appear despite the fact 
that our model assumes constant order parameter amplitude and the effective
core size is the tight binding lattice spacing $\delta$. The small size
of the core causes the lowest bound state to be pushed to rather high energy
and also that only few bound states can be resolved with our numerical 
accuracy. For energies 
$\epsilon\gg {\Delta}_0$ the spectrum exhibits Landau level quantization,
as expected from the argument presented above.

It is appropriate to illustrate the pitfall lurking in guise of the symmetric 
transformation widely used in the literature. At the first glance, perhaps 
the most natural choice  
for removing the phases from the off-diagonal terms is setting in Eq. 
(\ref{con1}) ${\phi}_e ({\bf r})={\phi}_h ({\bf r})={\phi}({\bf r})/2$. 
Note that in this case the transformation
\begin{equation}
U= \left( \begin{array}{cc}
 e^{i {\phi ({\bf r})/2}} & 0 \\
 0 & e^{-i {\phi ({\bf r})/2}}
\end{array} \right)  
\end{equation}
is not single valued and neither are the resulting wavefunctions. 
Nevertheless, ignoring these facts, the Hamiltonian ${\cal H}_N$ becomes
\begin{eqnarray}
\left( \begin{array}{cc}
-t \sum_{{\bf \delta}}e^{i {\cal V}_{\bf\delta}(\br)}
\hat{s}_{{\bf \delta}}- \epsilon_F &
 \Delta_{0} \\
 \Delta_{0} & t \sum_{{\bf \delta}}e^{-i {\cal V}_{\bf\delta}(\br)}
\hat{s}_{{\bf \delta}}+ \epsilon_F
\end{array} \right)
\label{u3}
\end{eqnarray}
with 
\begin{equation}
{\cal V}_{\bf\delta}(\br)=\int_{{\bf r}}^{{\bf r} + 
{\bf \delta}}({1\over 2}\nabla \phi - 
\frac{e}{\hbar c} {\bf A})\cdot d{\bf l}~.
\end{equation}
In the limit of high quasiparticle energies the equations again decouple, 
but now they 
describe a quasiparticle moving in a uniform magnetic field pierced by
half electronic flux quanta ${h c}/{2e}$ canceling the overall field. 
These half-fluxes cause significant Aharonov-Bohm 
scattering and cannot be ignored. As shown in Ref. \cite{nielsen}, the 
spectrum for this problem is {\em not} that of Landau levels;
there is a significant dispersion. We see that symmetric transformation
leads to the results that are manifestly incorrect. Again, this argument is 
independent of the pairing symmetry.


\subsection{$p$-wave pairing}

\begin{figure}
\epsfxsize=8.5cm
\hfil\epsfbox{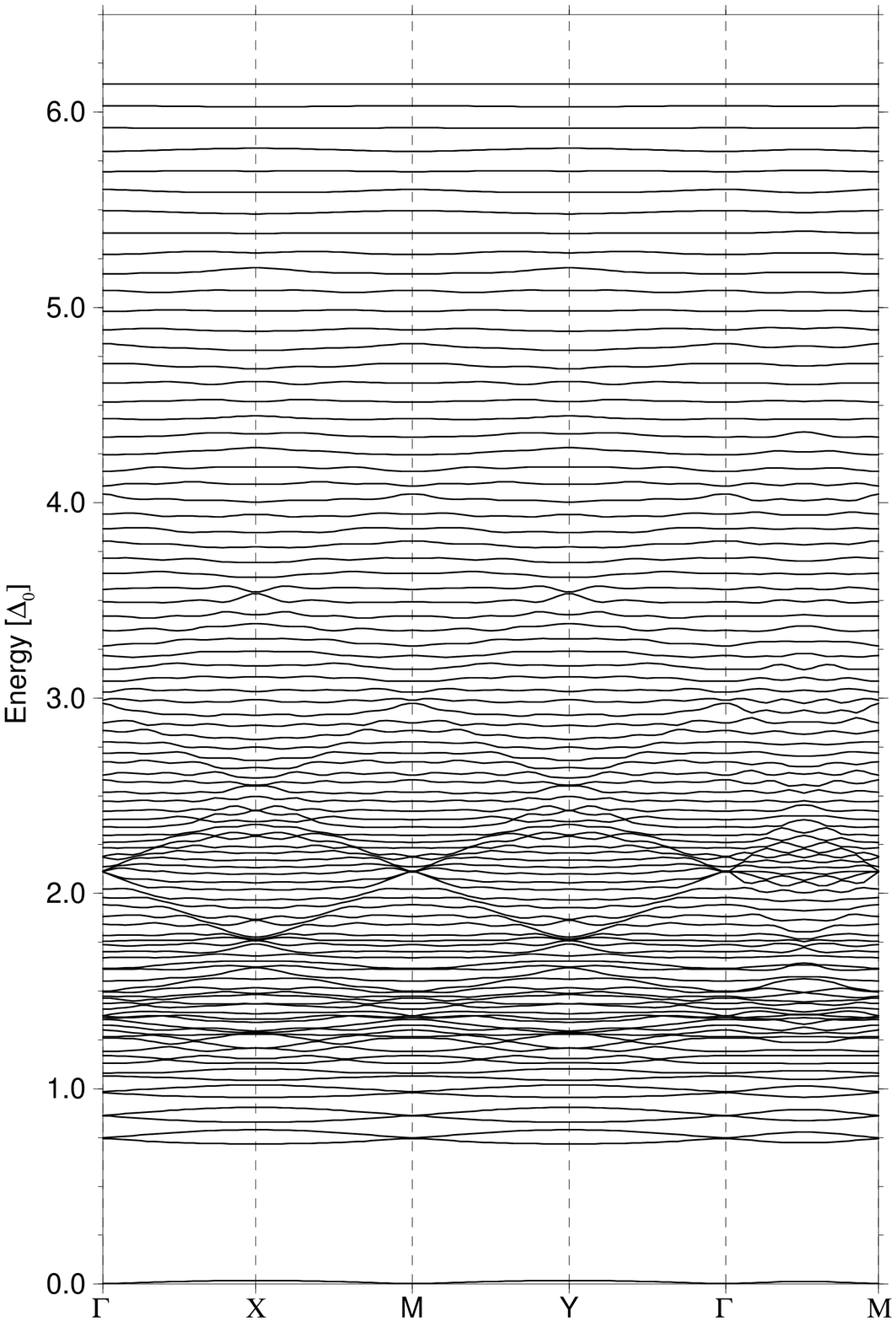}\hfill
\vspace{0.5cm} 
\epsfxsize=8.5cm
\hfil\epsfbox{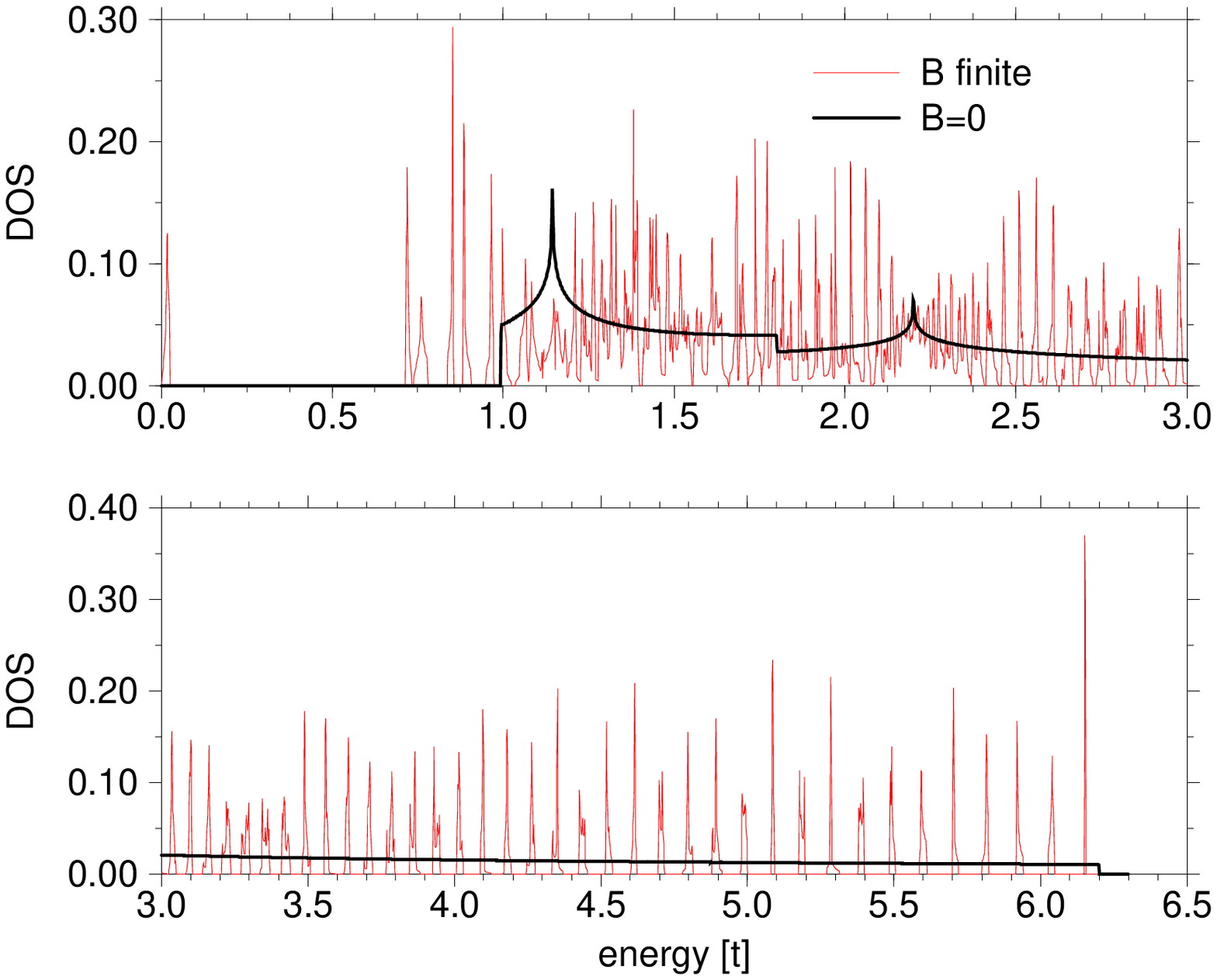}\hfill
\caption{Top: Quasiparticle band spectrum for a $p_{x+iy}$-wave superconductor 
in the presence of the external magnetic field B=1/100$\delta^2$, and  
$\Delta_0=t$, $\epsilon_F=-2.2t$. Bottom: The corresponding DOS. }
\label{pwaveband}
\end{figure}

We follow Matsumoto and Sigrist \cite{Matsumoto99} and for simplicity assume
that the prototype $p$-wave superconductor is two dimensional and has a 
cylindrical Fermi surface.
We further assume for simplicity that it is strongly type-II, although 
Sr$_{2}$RuO$_{4}$ is not of this type. 
We restrict the Hamiltonian to one of the two degenerate states, 
$p_{x+iy}$ and ignore the $p_{x-iy}$ part.  
For $p$-wave $(p_{x}+i p_{y})$ we have
\begin{equation}
 \hat{\eta}_{{\bf \delta}}= \left\{ \begin{array}{ll}
 \mp i \hat{s}_{{\bf \delta}} & \mbox{if $ {\bf \delta}=\pm \hat{x}$ } \\
 \pm \hat{s}_{{\bf \delta}} & \mbox{if $ {\bf \delta}=\pm \hat{y}$}  \end{array} \right. 
\end{equation}
where $\hat{s}_{{\bf \delta}} \,u({\bf r}) =u({\bf r}+{\bf \delta})$.
The Hamiltonian becomes
\begin{eqnarray}
\left( \begin{array}{cc}
-t \sum_{{\bf \delta}}e^{i {\cal V}_{\bf\delta}^A(\br)}
\hat{s}_{{\bf \delta}}- \epsilon_F &
 \Delta_{0} \sum_{{\bf \delta}}e^{i{\cal A}_{\bf \delta}(\br)} 
\hat{\eta}_{{\bf \delta}}\\
 \Delta_{0} \sum_{{\bf \delta}}e^{-i{\cal A}_{\bf \delta}(\br)} 
\hat{\eta}_{{\bf \delta}} &
t \sum_{{\bf \delta}}e^{-i {\cal V}_{\bf\delta}^B(\br)}
\hat{s}_{{\bf \delta}}+ \epsilon_F
\end{array} \right)
\label{hp}
\end{eqnarray}
where the phases ${\cal V}_{\bf\delta}^\mu(\br)$ are defined by Eq.\ 
(\ref{calv}) and 
\begin{equation}
{\cal A}_{\bf\delta}(\br)={1\over 2}\int_{{\bf r}}^{{\bf r} + 
{\bf \delta}}(\nabla \phi_A - \nabla \phi_B)\cdot d{\bf l}~.
\label{cala}
\end{equation}

We chose  $\epsilon_F=-2.2t$ which yields approximately circular Fermi surface
with the superconducting gap, to a good accuracy, uniform everywhere on 
the Fermi surface. As in the case of $s$-wave, the value of $\Delta_0$ is 
set equal to $t$. 
The resulting spectrum and density of states are shown in the Fig.\
(\ref{pwaveband}).

The spectrum again reveals bound vortex states broadened into a band. In 
contrast to the $s$-wave case we now have a state at zero energy. These 
results are what is expected on the basis of our understanding
of a single $p$-wave vortex \cite{Volovikpw99,Matsumoto99}.


\subsection{$d$-wave pairing}
\begin{figure}[t]
\epsfxsize=8.5cm
\hfil\epsfbox{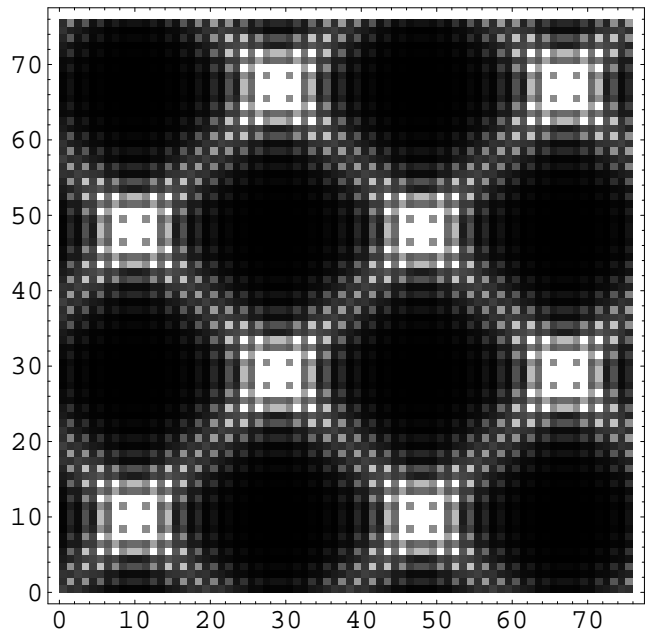}\hfill
\caption{
Local density of states at $E=0$ for the $d_{x^2-y^2}$-wave superconductor 
with square arrangement of vortices. The plot is in units of the tight 
binding lattice constant $\delta$. Bright regions represent maxima 
while the dark regions represent minima. 
The parameters are $\epsilon_F =0$, $\alpha_D=t/\Delta_0=4$.
The strong overlaps of the low energy 
quasiparticle wave functions along the four nodal directions cause appreciable
interference effects which in turn influence the character of the 
quasiparticle spectrum. 
}
\label{dwaveSQinterf}
\end{figure}

To model the high-temperature superconductors such as YBa$_2$Cu$_3$O$_7$, 
we assume that coupling between the 
Cu-O planes is weak and to the leading approximation can be ignored. 
On the tight binding lattice the $d$-wave pairing is given by 
$\Delta = 2 \Delta_0 [\cos(k_x \delta) -\cos(k_y \delta)] $ which determines 
the form of the lattice operator to be: 
\begin{equation}
 \hat{\eta}_{{\bf \delta}}= \left\{ \begin{array}{ll}
     \hat{s}_{{\bf \delta}} & \mbox{if $ \bf{ \delta } = \pm \hat{x}$ } \\
    -\hat{s}_{{\bf \delta}} & \mbox{if $ \bf{ \delta } = \pm \hat{y}$}  
\end{array} \right. 
\end{equation}
where as before $\hat{s}_{{\bf \delta}} \,u({\bf r}) =u({\bf r}+{\bf \delta})$.
With this definition of $\hat{\eta}_{{\bf \delta}}$ the Hamiltonian for 
$d$-wave pairing has the same form as Eq.\ (\ref{hp}).
\begin{figure}[t]
\epsfxsize=8.5cm
\hfil\epsfbox{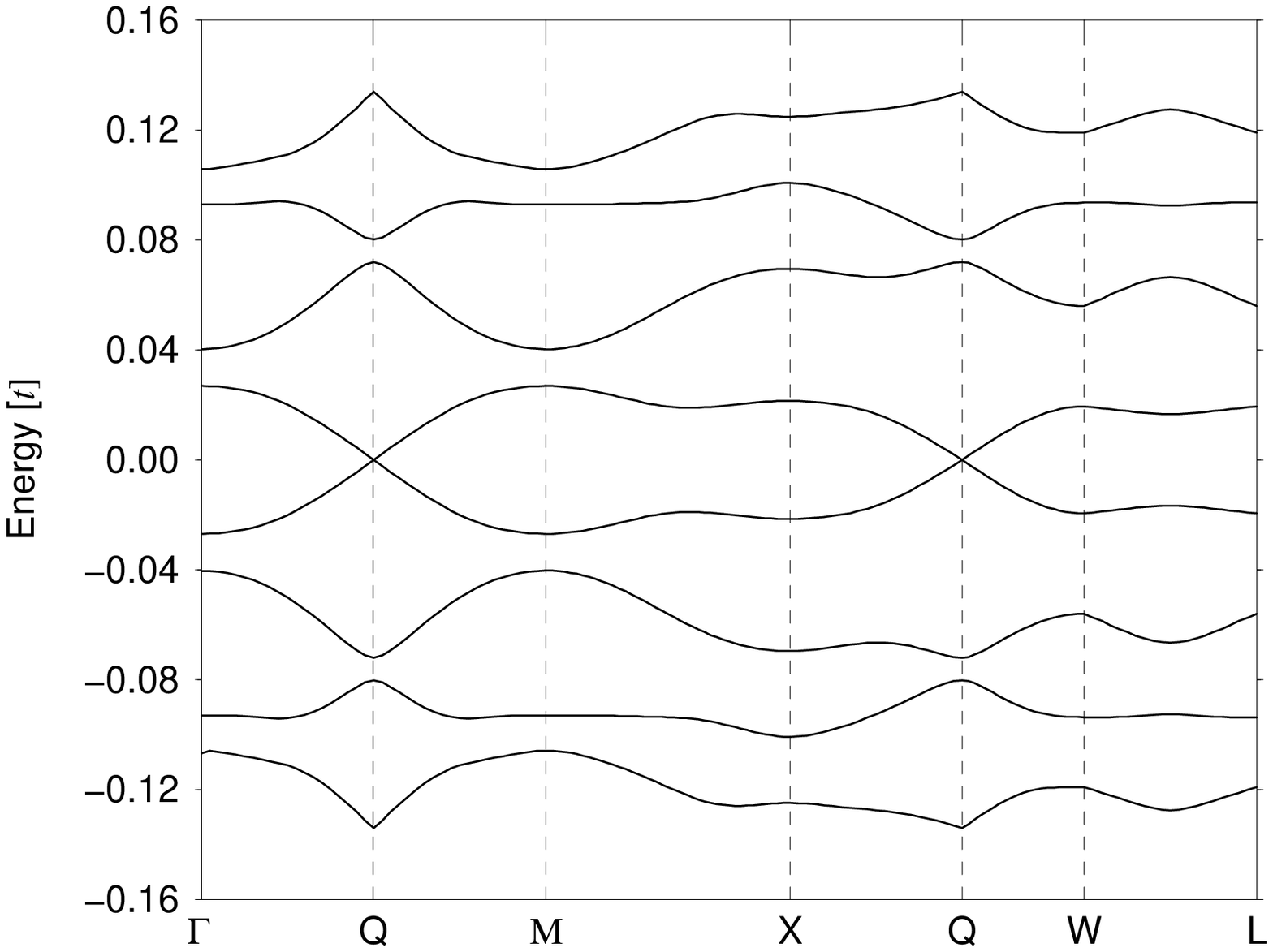}\hfill
\vspace{.5cm}
\epsfxsize=8.5cm
\hfil\epsfbox{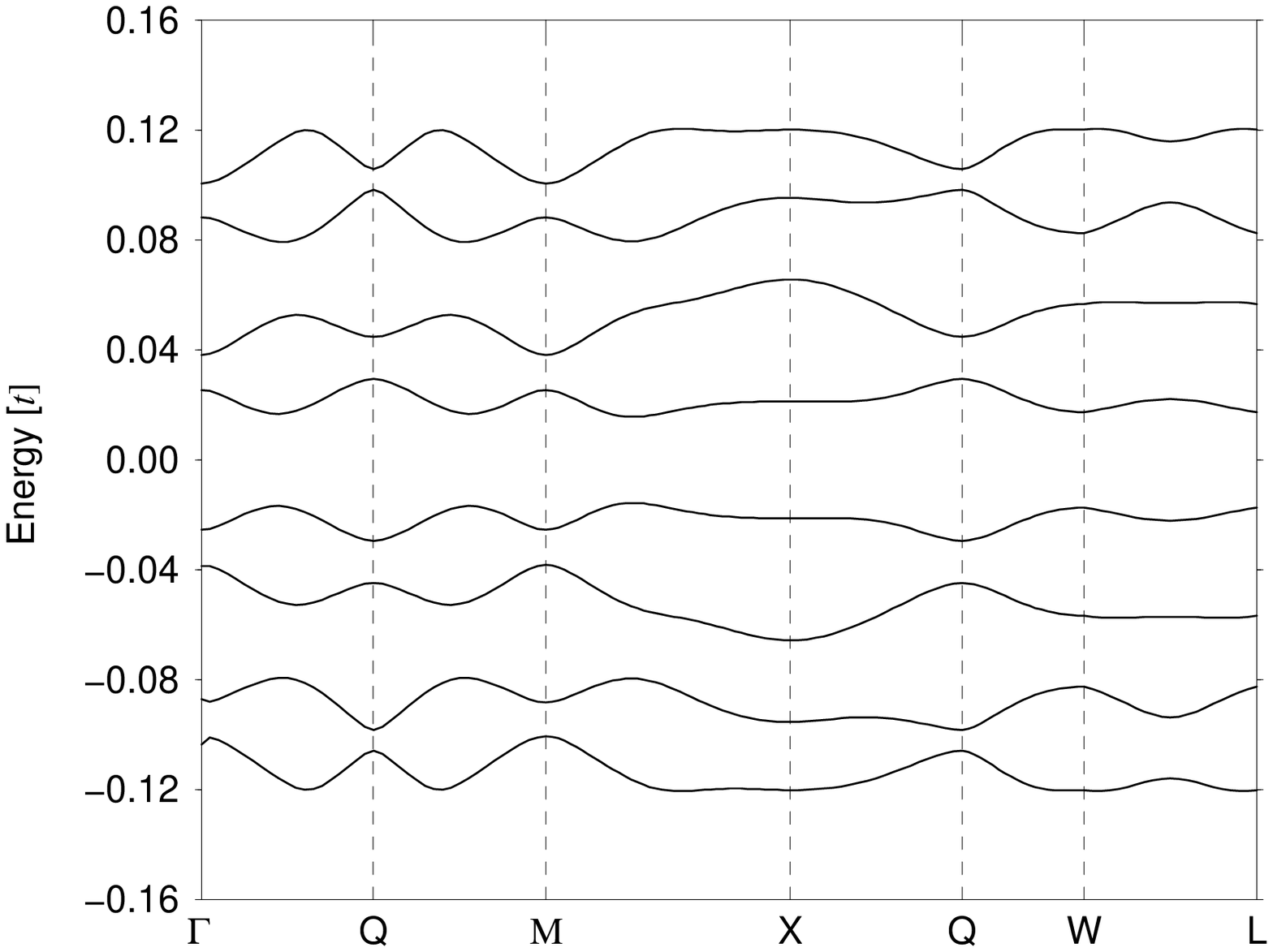}\hfill
\caption{Low energy part of the quasiparticle
band spectrum for the square vortex lattice.
The parameters are $\epsilon_F =0$, $\alpha_D=4$.
Top:  example of a gapless spectrum for $l=38\delta$.
Note that the node at $Q$ is moved away from the original
Dirac $\Gamma$ point. This effect is a result of a uniform gauge 
``boost'' associated with the choice of vortex unit cell and disappears
for a unit cell with four vortices. Bottom: example of a gapped spectrum
with $l=40\delta$.}
\label{dwave38banda4}
\end{figure}

The results presented in this section correspond to
the magnetic field $\phi_0 /(1600 {\delta}^2 )$ 
for square vortex lattice
and $\phi_0 /(1500 {\delta}^2 )$ for triangular vortex lattice
where $\phi_0= hc/e=4.137\times 10^5$ T\AA$^2$ and $\delta$ is the tight 
binding lattice constant. Taking $\delta=4$ \AA, as in YBa$_2$Cu$_3$O$_7$, this
corresponds to physical field of 16 T.
The above parameters were chosen for computational efficiency, but
we did not see any qualitative difference down to
the fields as low as $\phi_0 /(4900 {\delta}^2 )$ corresponding to a 
magnetic unit cell of 70$\delta\times70\delta$ and a field of 5.2 T. 
Numerical diagonalization was performed using the 
ARPACK package routines for sparse matrices. This algorithm
provides a set of low-lying eigenvalues and allows handling  
much larger systems than the full diagonalization used in $s$- and $p$-wave
cases. 

We find that the quasiparticle wave functions
exhibit significant dependence on the symmetry of the vortex lattice. 
For the square lattice, the overlap among
the wave functions corresponding to different nodes 
is appreciable and there are strong interference effects 
along the $|x|=|y|$ diagonals, i.e. the directions in the real space 
where $\Delta({\bk})$ vanishes. This is illustrated in 
Fig.~\ref{dwaveSQinterf}. For certain commensurability of the tight binding 
and the square vortex  lattices, the 
interference effects are responsible for {\em opening a gap} at Fermi energy,
while for the complementary set of lattices at different commensurability 
factors, the spectra are gapless at the Dirac $\Gamma$ point. 

\begin{figure}
\epsfxsize=8.5cm
\hfil\epsfbox{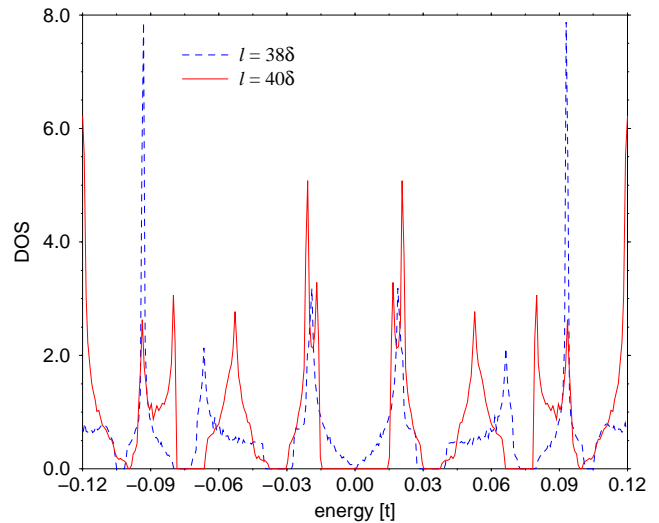}\hfill
\caption{
~Low energy part of the
quasiparticle density of states for an $d_{x^2-y^2}$-wave superconductor 
with square
 arrangement of vortices for two different interference cases in which
${\bf k}_F \cdot  {\bf d} = 2n  \pi \;$ (solid line)
and ${\bf k}_F \cdot{\bf d} = (2n+1) \pi \;$ (dashed line), $n$ being an
integer, ${\bf k}_F=(\frac{\pi}{2},\frac{\pi}{2})$ a Fermi vector at
nodal points,
and ${\bf d}$ is the primitive vortex lattice vector. 
Notice the appearance of the gap in the density of states for ${\bk_F}
\cdot{\bf d} = 2n  \pi $.  
Plotted on arbitrary scale, energy is in units of $t$. 
The parameters are $\epsilon_F =0$, $\alpha_D=4$.
}
\label{dwaveSQ}
\end{figure}

\begin{figure}[t]
\epsfxsize=8.5cm
\hfil\epsfbox{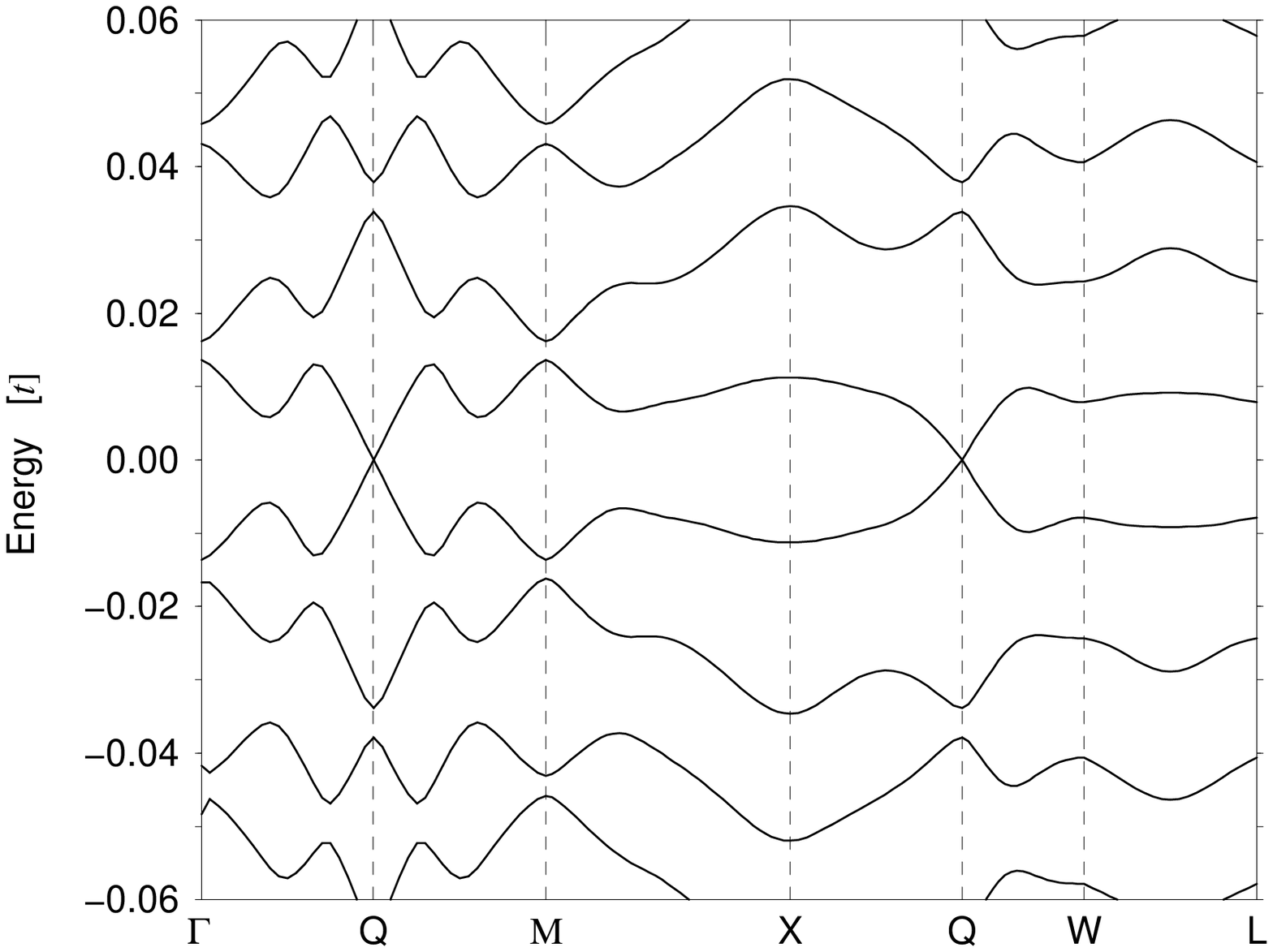}\hfill
\vspace{.5cm}
\epsfxsize=8.5cm
\hfil\epsfbox{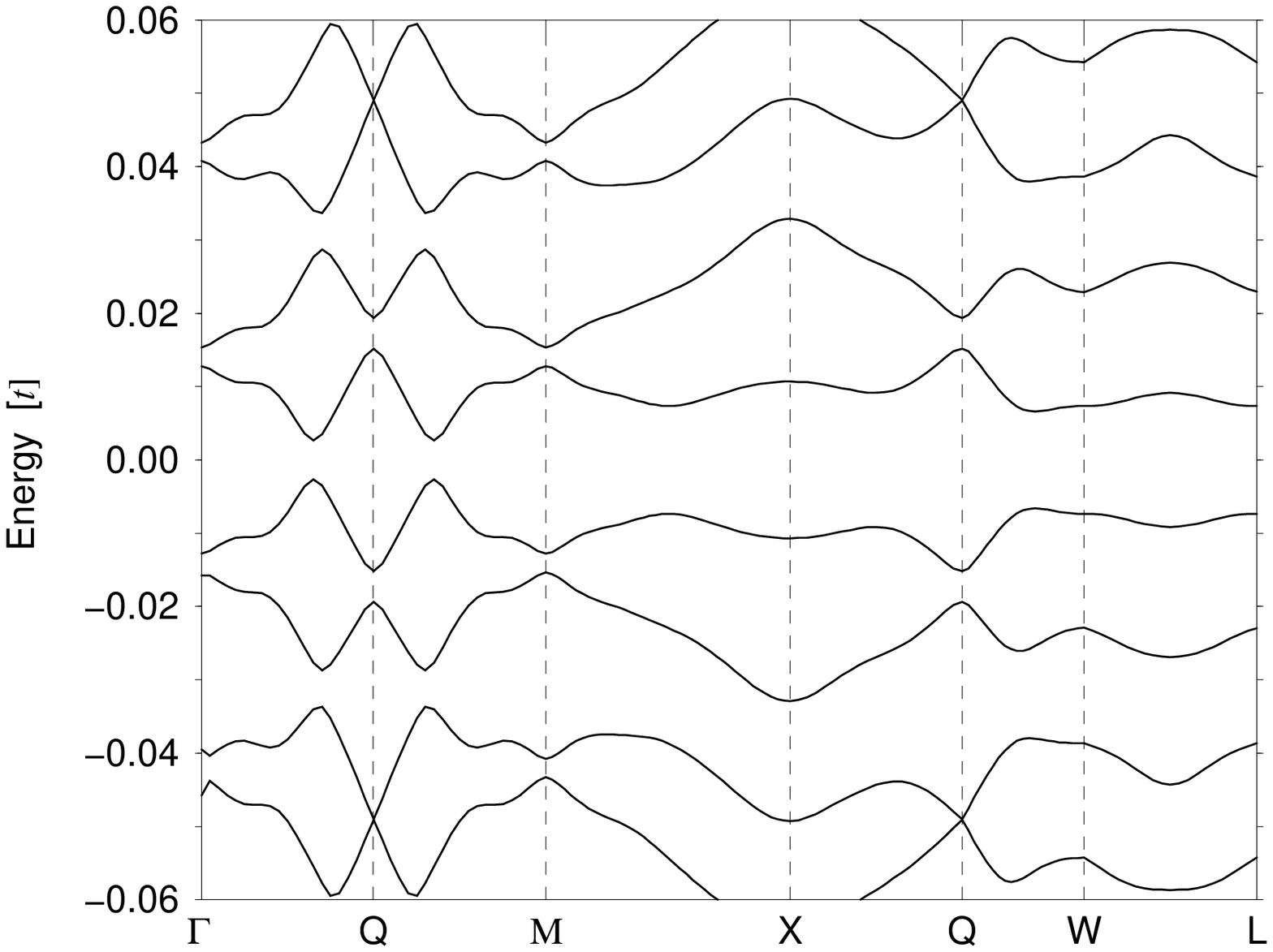}\hfill
\caption{
Low energy part of the quasiparticle
band spectrum for the square vortex lattice.
The parameters are $\epsilon_F =0$, $\alpha_D=10$. 
Top: gapless spectrum for  $l=38\delta$.
Note the increase of the dispersion in the
QM direction with increase of the Dirac anisotropy $\alpha_D$.
Bottom: gapped spectrum for $l=40\delta$.
}
\label{dwave38banda10}
\end{figure}

\begin{figure}
\epsfxsize=8.5cm
\hfil\epsfbox{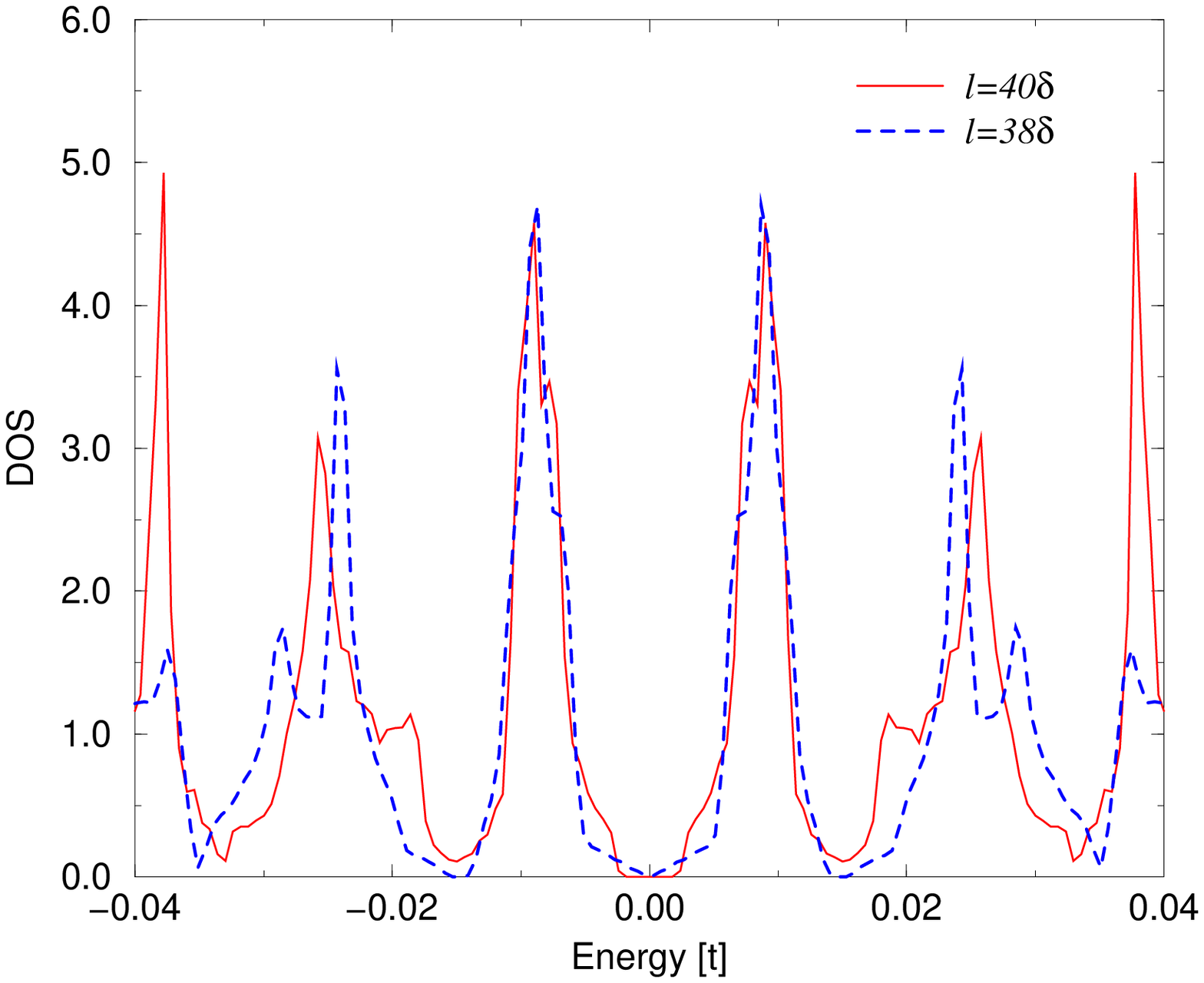}\hfill
\caption{
~Low energy part of the
quasiparticle density of states for an $d_{x^2-y^2}$-wave superconductor 
with square 
arrangement of vortices for two different interference cases: $l=38\delta$ (dashed)
and $l=40\delta$ (solid).
Plotted on arbitrary scale, energy is in units of $t$. 
The parameters are $\epsilon_F =0$, $\alpha_D=10$.}
\label{dwaveSQ10}
\end{figure}

\begin{figure}[t]
\epsfxsize=8.5cm
\hfil\epsfbox{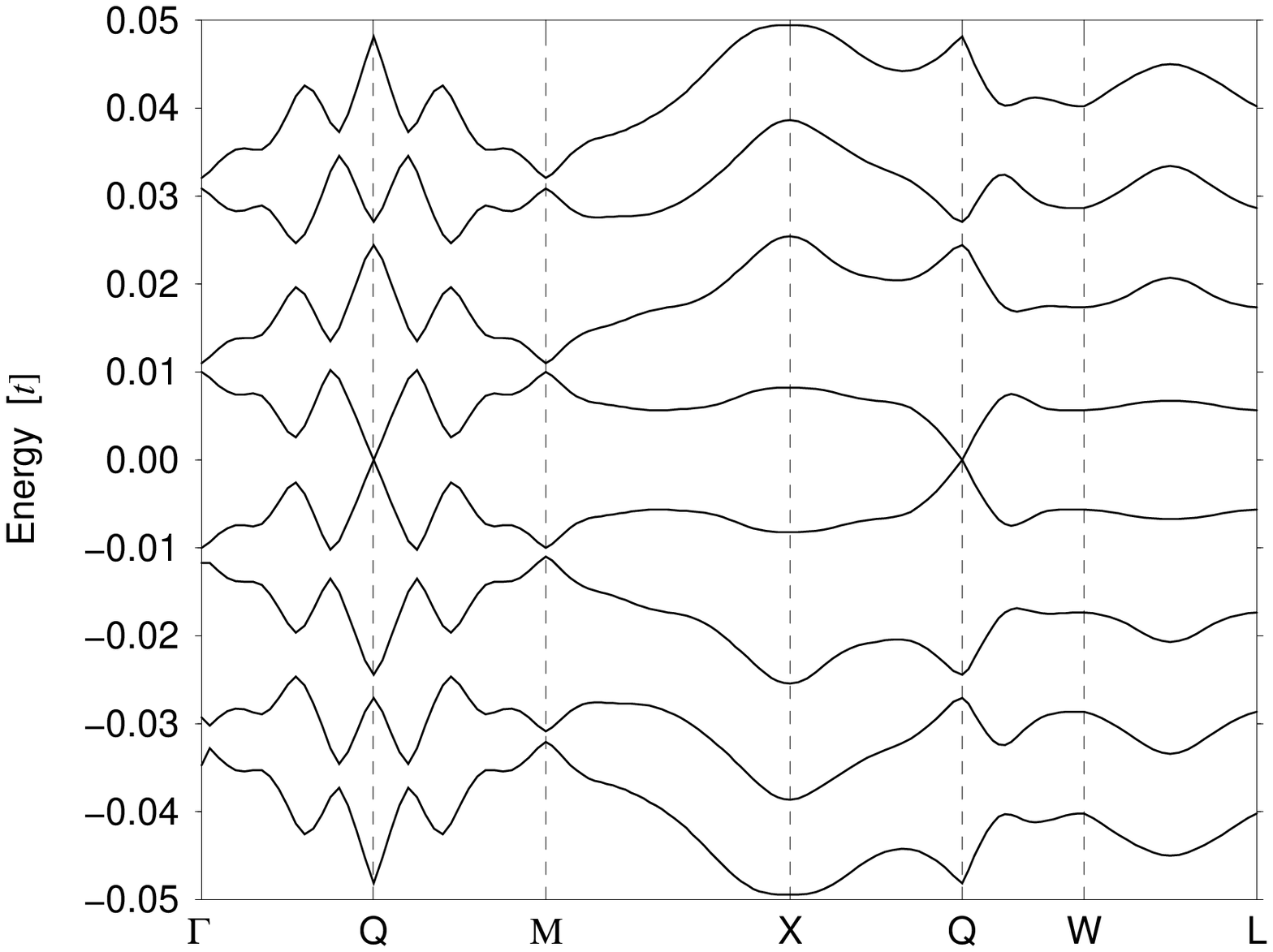}\hfill
\vspace{.5cm}
\epsfxsize=8.5cm
\hfil\epsfbox{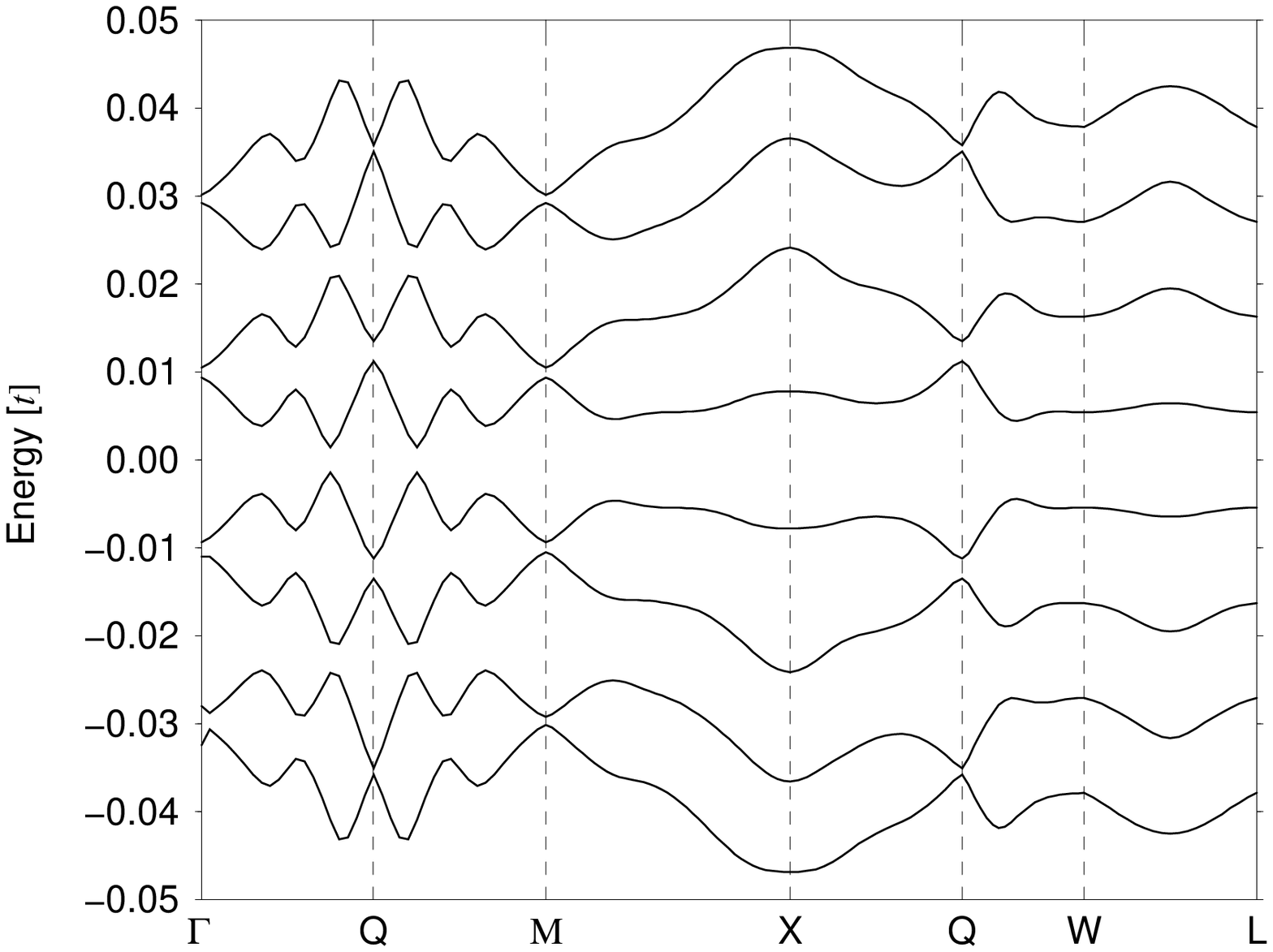}\hfill
\caption{Low energy part of the quasiparticle
band spectrum for the square vortex lattice.
The parameters are $\epsilon_F =0$, $\alpha_D=15$. 
Top: gapless spectrum for  $l=38\delta$.
Note the increase of the dispersion in the
QM direction with increase of the Dirac anisotropy $\alpha_D$.
Bottom: gapped spectrum for $l=40\delta$.
}
\label{dwave38banda15}
\end{figure}

\begin{figure}
\epsfxsize=8.5cm
\hfil\epsfbox{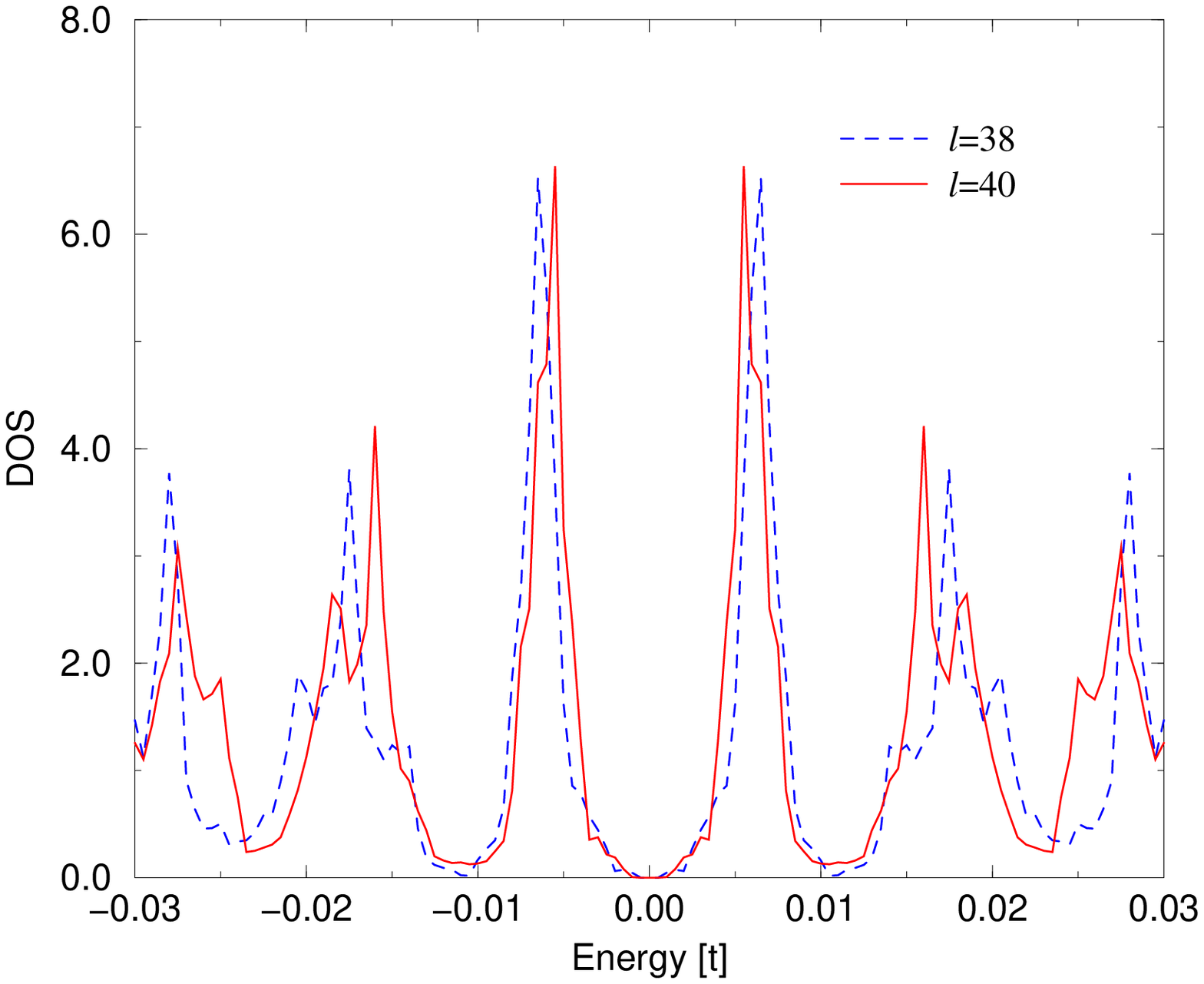}\hfill
\caption{
~Low energy part of the
quasiparticle density of states for an $d_{x^2-y^2}$-wave superconductor with square
 arrangement of vortices for two different interference cases: $l=38\delta$ (dashed)
and $l=40\delta$ (solid).
Plotted on arbitrary scale, energy is in units of $t$. 
The parameters are $\epsilon_F =0$, $\alpha_D=15$.}
\label{dwaveSQ15}
\end{figure}

Figs.~\ref{dwave38banda4}
and \ref{dwaveSQ} show the low-energy band structures and
the low-energy density of states for the square
lattice. The two system sizes shown illustrate the commensurability effect:
if the scalar product between the Fermi vector along the nodal direction ${\bf k}_F$ and the vortex
primitive Bravais lattice vector ${\bf d}$ is an even integer times $\pi$,
the spectrum develops a gap, while it remains gapless if this product is an
odd integer times $\pi$. The same effect is seen at higher Dirac
anisotropy $\alpha_D\equiv t/\Delta_0 = 10$ (Figs.~\ref{dwave38banda10}, \ref{dwaveSQ10}) and $\alpha_D=15$  
(Figs.~\ref{dwave38banda15}, \ref{dwaveSQ15}).

%

%


\begin{figure}
\epsfxsize=8.5cm
\hfil\epsfbox{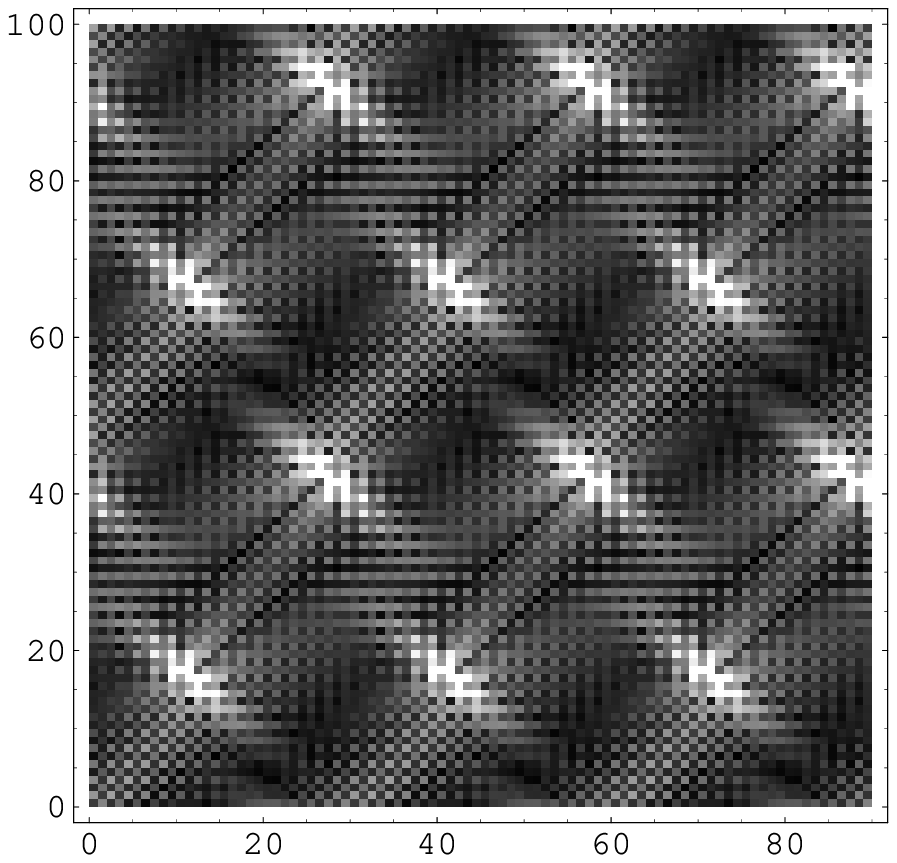}\hfill
\caption{
Displayed is a typical low energy modulus of a quasiparticle 
wavefunction for the 
$d_{x^2-y^2}$-wave superconductor with triangular arrangement of vortices. 
The plot is in units of the tight binding lattice constant $\delta$. 
Bright regions represent maxima 
while the dark regions represent minima. 
The parameters are $\epsilon_F =0$, $\alpha_D=4$.
This plot illustrates the reduction of the interference effects for
triangular vortex lattice. As a
consequence the gap in the quasiparticle spectrum does not emerge as shown in
Fig.~\ref{dwaveTR}.}
\label{dwaveTRIinterf}
\end{figure}

\begin{figure}
\epsfxsize=8.5cm
\hfil\epsfbox{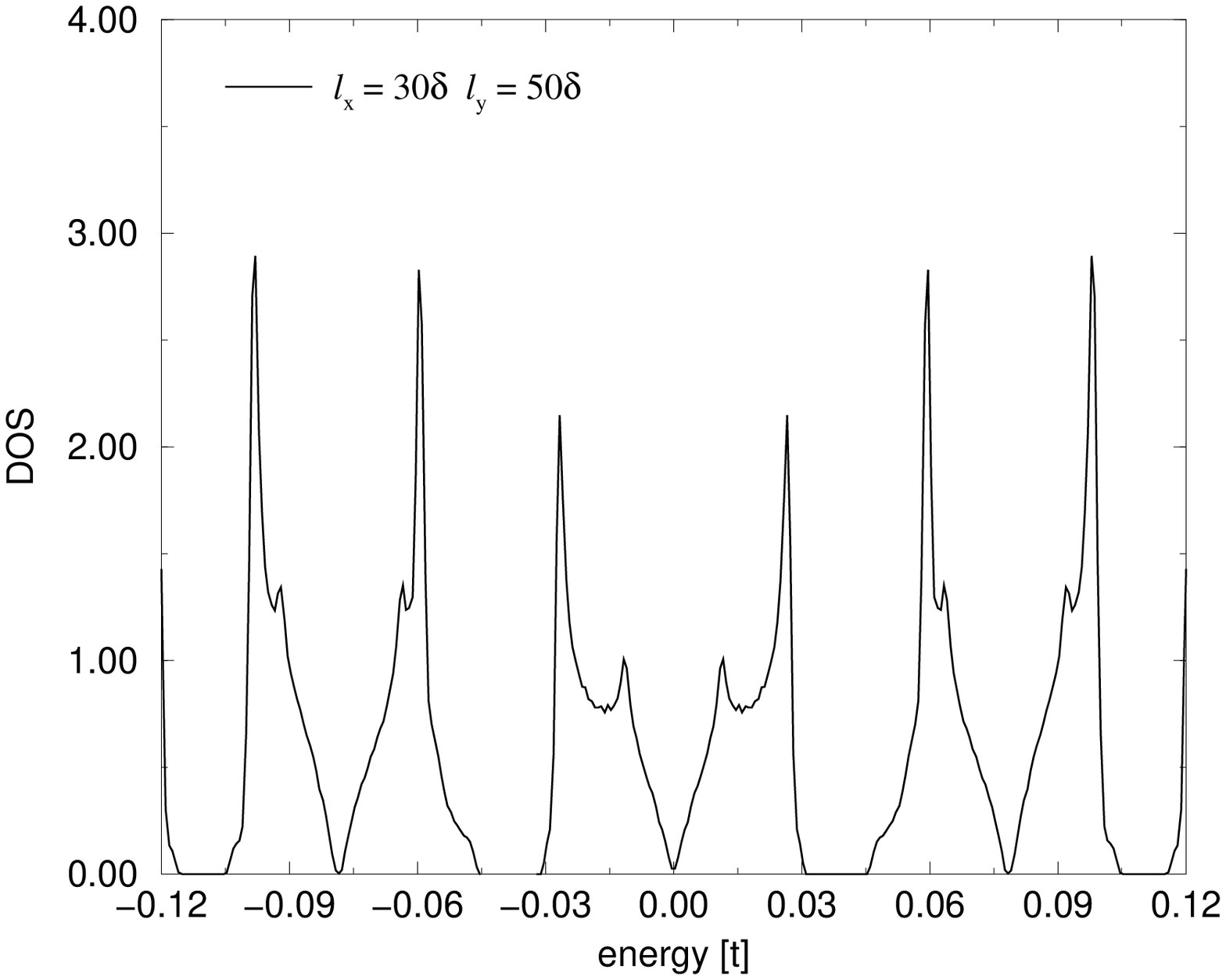}\hfill
\caption{Low energy part of the 
quasiparticle density of states for an $d$-wave superconductor with triangular
arrangement of vortices. Plotted on arbitrary scale, energy is in units of 
$t$. The parameters are $\epsilon_F =0$, $\alpha_D=4$.}
\label{dwaveTR}
\end{figure}

These interference effects persist down to low magnetic fields where the 
interference gaps scale as $\sim\sqrt{B}$ (see the next section and 
Fig.~\ref{gapscaling}).
In the case of a triangular lattice, the interference effects were greatly 
reduced (Fig.~\ref{dwaveTRIinterf}) and no commensurability dependence was 
observed. We find the spectrum to be gapless at half filling in this case 
(see Fig.~\ref{dwaveTR}).

Finally we note that we have explicitly verified that identical spectra
(to within numerical accuracy) are found irrespective of our assignment of 
the $A$-$B$ sublattices. This finding confirms that the choice of  
$A$-$B$ vortices is an internal gauge symmetry of the problem, as one would
expect on general grounds.


\section{Discussion}

\subsection{Comparison of continuum and lattice results}
The results of Sections II and III show that, under generic conditions, 
the spectrum of linearized Dirac Hamiltonian provides a reasonable 
approximation to the low energy part of the spectrum of the 
full BdG Hamiltonian.
The Dirac nodes are preserved provided that there is no commensuration
between $\bk_F$ at the node and the primitive vector of the vortex lattice
${\bf d}$, and thus the internodal scattering can be neglected.
The overall shape of the energy bands is also qualitatively  and 
quantitatively similar for the two cases. 

Previous investigations of the linearized Hamiltonian\cite{franz1,marinelli}
established that the spectrum becomes quasi one dimensional and lines of nodes
appear\cite{remark3} 
in the Brillouin zone for large Dirac cone anisotropy $\alpha_D>14$,
leading to finite DOS at the Fermi level. Inspection of Fig.\ 
\ref{dwave38banda15} suggests that similar effect takes place in the full
BdG Hamiltonian, although the one dimensionality is somewhat less pronounced
and is restricted to the immediate vicinity of the node. Furthermore, the
lines of nodes never quite form (although the tendency is clearly visible
along the line $\Gamma\to M$) and the DOS remains zero at the Fermi level.  
In the above discussion one needs to bear in mind that the vortex lattice 
considered here has
been rotated by $45^{\circ}$ relative to Ref.\cite{franz1}.


\subsection{Scaling of the Energy Spectrum}
\begin{figure}
\epsfxsize=8.5cm
\hfil\epsfbox{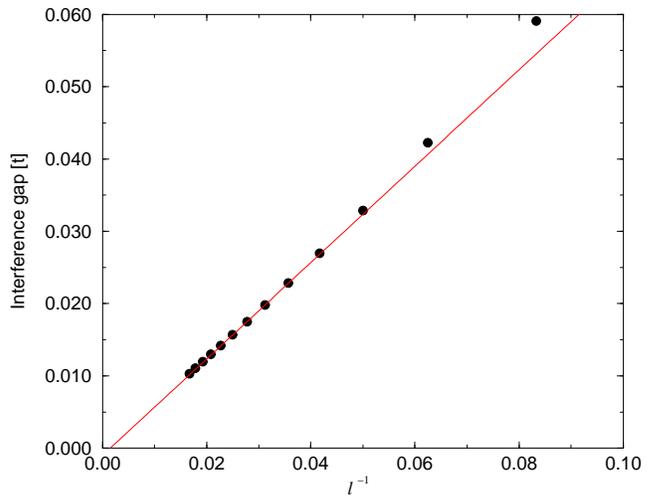}\hfill
\caption{
The magnitude of the interference gaps vs. magnetic length $l$ exhibits
$l^{-1}$ scaling. $\Delta_0=0.25t$, $\epsilon_F=0$.
}
\label{gapscaling}
\end{figure}

The wavefunction interference effects among the four nodes, which are
responsible for opening of the ``interference" gaps visible in some
of our spectra, seem to be in contrast with the results obtained for the 
linearized Hamiltonian by FT \cite{franz1} and more
recently by Marinelli, Halperin, and Simon \cite{marinelli}, where
any internodal interaction is ignored and assumed insignificant. Furthermore, 
Marinelli {\em et al.} advanced strong analytic arguments that in the presence 
of particle-hole symmetry the linearized Hamiltonian retains the
Dirac node at the $\Gamma$ point and does not develop a gap to order 
${\cal O}\left(l^{-1}\right)$, where $l$ is the magnetic length. 
We found that the ``interference"  gaps in the quasiparticle spectrum 
in the case of the square lattice scale with magnetic field as 
$\sqrt{B} \sim l^{-1}$. This is shown in Fig.~\ref{gapscaling}. 
The reason for this can be understood from the scaling of the wavefunctions 
of the 
linearized Hamiltonian. We find that there is a $r^{-1/2}$ divergence in the 
asymptotic
solution of the wavefunction around one vortex. This strong concentration of 
the wavefunction 
around the vortex makes the contribution from the term quadratic in the 
superfluid velocity 
particularly enhanced and, independent of regularizing the wavefunctions to 
eliminate the 
divergences at the core, the contribution to the gap is significant. We can
then extract the 
dependence of the wavefunction on the magnetic length $l$ just from 
dimensional analysis. The
wavefunction must have units of length$^{-1}$ therefore 
$\psi \sim (rl)^{-1/2}$. One can see, that
the matrix element, and consequently the gaps,
will in general scale as $l^{-1}$ for the terms beyond the 
linearized Hamiltonian. This dependence is extremely difficult to obtain from 
the plane
wave expansion of the wavefunctions.

Our numerical results strongly suggest that there is a characteristic 
oscillation in the
gap of the spectrum depending on the commensurability of the magnetic lattice 
and the underlying ionic lattice. This can be interpreted as the inter-nodal 
scattering. The interaction
between the quasiparticles at different nodes is responsible for opening the 
gaps at the
Fermi surface. The effect of the intra-nodal scattering on these gaps on the 
Fermi surface 
is, however, absent since for certain commensurability of the magnetic and 
ionic lattices 
there is no gap. Thus we conclude that the effect is purely due to the 
internodal scattering mediated by the terms beyond the linearized Hamiltonian.
 The sensitivity of the gaps to the commensurability of the ionic and magnetic 
lattices is supported by the results with triangular vortex lattice, in which 
case the spectrum remains gapless
as there is no commensurability between ionic and vortex lattice.
This supports the view that the internodal scatting alone is responsible for 
the presence of the gap at the Fermi surface.  


\subsection{Comparison with the Doppler-Shift-Only Results for $d$-wave gap}

One of the key insights gained from the FT 
transformation is that the familiar and often used
Doppler shift approximation \cite{Volovikdw93,sauls} is
not sufficient to describe the quasiparticle dynamics in the vortex lattice. 
While the Doppler shift enters at the level of a linearized
Dirac Hamiltonian as a periodic {\it scalar} potential, there is also 
an effective {\it vector} potential {\bf a} (Sec. II), 
which originates from the global
curvature of the superflow in the presence of vortices. This new vector
potential term leads to additional strong magnetic half-flux scattering 
across the whole energy spectrum.
It is instructive to compare our results with those obtained by 
performing the symmetric gauge transformation specified by Eq.\ (\ref{u3}).
As already pointed out the symmetric transformation is {\it not} 
single-valued and the resulting 
transformed Hamiltonian must be accompanied by the
branch cuts imposed on its eigenfunctions. If one simply {\it ignores}
these branch cuts altogether, the resulting Hamiltonian  ${\cal H}_S$ contains
only the Doppler shift terms and reads 
\begin{eqnarray}
\left( \begin{array}{cc}
-t \sum_{{\bf \delta}}e^{i {\cal V}_{\bf\delta}(\br)}
\hat{s}_{{\bf \delta}}- \epsilon_F &
 \Delta_{0} \sum_{{\bf \delta}} \hat{\eta}_{{\bf \delta}}\\
 \Delta_{0} \sum_{{\bf \delta}}\hat{\eta}_{{\bf \delta}} &
t \sum_{{\bf \delta}}e^{-i {\cal V}_{\bf\delta}(\br)}
\hat{s}_{{\bf \delta}}+ \epsilon_F
\end{array} \right).
\label{hds}
\end{eqnarray}

\begin{figure}
\epsfxsize=8.5cm
\hfil\epsffile{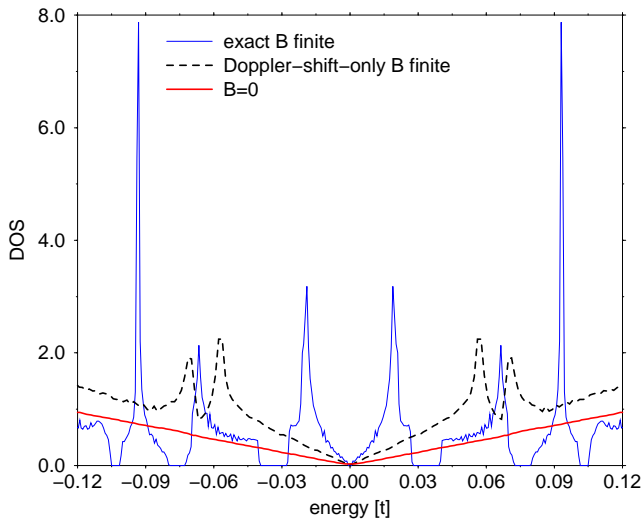}\hfill
\caption{
Comparison of the low energy part of the 
quasiparticle density of states for an $d$-wave superconductor with square
arrangement of vortices with the DOS obtained from the Doppler-shift-only
approximation.  Plotted on arbitrary scale, the energy is in units of $t$. 
The parameters are $\epsilon_F =0$, $\l=38\delta$, $\alpha_D =4$.
}
\label{volovik}
\end{figure}

The density of states
obtained by diagonalizing ${\cal H}_S$ is shown Fig.~\ref{volovik}.
It is significantly different from the results presented in Section III. 
This clearly demonstrates 
that there is an essential piece of physics missing from 
the Hamiltonian which contains only the Doppler shift effect, 
and consequently from the frequently encountered semiclassical 
approximation to such Hamiltonian \cite{Volovikdw93}.


\section{Conclusions}

In conclusion, general utility of the singular gauge transformation
for the calculation of the quasiparticle spectra in the vortex state of
a general pairing symmetry was shown. Once the tight binding regularization 
is introduced,
the spectrum can be computed in principle exactly using the Bloch states as
the natural basis, although one is bound to resort to numerical calculation 
regardless of
respecting the self-consistency. In the case of $s$- and $p$-wave 
symmetry we showed
that the method applied to an array of vortices leads to results 
consistent with single vortex solutions.
	
For $d$-wave pairing spectrum is also consistent with the single 
vortex solution from the point of view that the all the quasiparticle
wavefunctions are delocalized 
and no bound states are observed. Additional insight is gained from the exact 
solution with respect to the continuum linearized version of the theory. 
For specific commensurability of the tight binding and square vortex lattice
the internodal scattering mediated by the terms neglected in the linearized 
theory is found to be significant and of the same order of magnitude 
as the terms present in the linearized Hamiltonian. 
This is believed to be brought forth by the diverging accumulation of the 
Dirac wavefunctions 
in the vicinity of the vortex core, consequently giving rise to increased 
significance of the terms beyond linearization. However, since rather 
special conditions must
be met for this effect to be significant, it is suggested that introduction 
of any perturbing agent such as disorder in the position of the vortices or 
vortex vibrations will lead to decoherence of the matrix elements 
for the internodal scattering and subsequently to the suppression of the
 interference effect. It is also possible, and in our view likely, that in a 
fully selfconsistent solution the vortex array would spontaneously undergo
a slight spatial deformation into an incommensurate state so as to
avoid opening gaps in the 
quasiparticle spectrum. In this respect, at chemical potential $\epsilon_F=0$, 
the results of the theory regularized on the 
tight binding lattice agree with the continuum linearized version.

We have uncovered a peculiar property of the {\em linearized}
 Dirac Hamiltonian:
it appears to violate the internal gauge symmetry associated with
the assignment of $A$ and $B$ vortices. Although the final resolution of this 
apparent contradiction awaits further research, we attribute it tentatively to
the unusually strong scattering of Aharonov-Bohm half fluxes acting in the 
Dirac Hamiltonian with unbounded excitation spectrum. This problem does not 
occur in the full BdG Hamiltonian. We conclude that the
original ``$ABAB$'' choice of the gauge\cite{franz1} is the one most 
representative
of the actual spectrum because it results in smoothest possible variation
of phase in the vortex lattice. This view is also supported by the direct
comparison with the spectrum obtained using the full BdG Hamiltonian. 

Number of intriguing issues remain to be addressed. In particular the effect
of static and dynamic disorder in the vortex positions on the quasiparticle
spectra must be understood in order to make connection with the
experimental data. Another set of unresolved 
issues arises in connection with the zero
field superconducting state phase-disordered by fluctuating vortex-antivortex
pairs\cite{franz98b,kwon99,balents}. One would expect that the ``Berry phase''
term arising from fluctuating 
vortices would influence in a profound way the critical
behavior of the HTS system on the verge of becoming a Mott insulator.

\acknowledgments

The authors are indebted to B. I. Halperin, L. Marinelli, 
N. Read, J. Ye and M. R. Zirnbauer for useful discussions.
This research was supported in part by NSF grant DMR-9415549.

\appendix
\section{Lattice and Continuum BdG Hamiltonian in a Gauge
Invariant Formulation}

In this Appendix we derive the explicit form of the pairing operator
$\hat\Delta$ (\ref{IIiii}) which appears in the lattice
BdG Hamiltonian of Eq. (\ref{IIi}). We also derive
the continuum limit of this operator which is used to construct
the BdG equations of Section II. Throughout our derivation we pay 
a special attention to the preservation of local gauge invariance.
We start with the general pairing term on a tight-binding square lattice:
\begin{equation}
\sum_{<i,j>}\Delta (i,j)\bigl [u^*(i)v(j)+u^*(j)v(i)\bigr ] + {\rm h.c.}
\label{ai}
\end{equation}
Here $\Delta (i,j)$ is a complex paring potential defined
on the nearest neighbor bonds. Such pairing term generically
arises when $t$-$J$  and related Hamiltonians, which are thought to represent
good microscopic models of various unconventional superconductors,
are treated within a BCS-type pairing approximation.
A conventional $s$-wave case follows from Eq. (\ref{ai}) if we replace
the sum over nearest neighbor bonds $<i,j>$ with the sum over
sites $i$. By construction, the pairing term (\ref{ai}) is
invariant under gauge transformations:
\begin{eqnarray}
u(i)&\to &  u(i)e^{i\chi(i)}, \nonumber \\
v(i) &\to & v(i)e^{-i\chi(i)},\\
\Delta(i,j) &\to & \Delta(i,j)e^{i\chi(i)+i\chi(j)}, \nonumber 
\end{eqnarray}
where $\chi(i)$ is an arbitrary non-singular function.

The actual form of the complex function
$\Delta(i,j)\equiv D(i,j)\exp(i\varphi (i,j))$ 
is obtained as a self-consistent solution of the
gap equation in some specific gauge. 
We denote its ``amplitude" and phase by $D(i,j)$
and $\varphi (i,j)$, respectively. For convenience, the ``amplitude"
$D(i,j)$ is defined so that it already contains the information
about the relative orbital state of a superconductor. For example,
in a pure $d_{x^2-y^2}$ superconductor and at zero field,
$D(i,j)=-(+) \Delta$ for $<i,j>$ in the $x(y)$ direction, where
$\Delta$ is a complex constant.
Furthermore, for the purposes of this paper, we assume that
the {\em actual} amplitude $|\Delta (i,j)|$ can be well approximated by a
uniform (real) constant $\Delta_0$, independent of $<i,j>$. This assumption is
valid in the space between vortices but it clearly
breaks down inside a vortex core. At low fields, 
$H_{c1}\ll H\ll H_{c2}$, where the intervortex 
separation is much larger than the core size we expect any
effect of the inhomogeneous amplitude to be negligibly small.

The essential information
about vortex configurations and self-consistent solution
at a finite magnetic field is now stored in the bond
phase $\varphi (i,j)$. Near a plaquette containing a vortex
$\varphi (i,j)$ changes rapidly from bond to bond. Far from
vortex cores, however, we expect $\varphi (i,j)$ to be some
smoothly varying function undergoing only small changes between
neighboring bonds. Consequently, in the regions far away from
vortex cores we can replace the {\em bond} phase $\varphi (i,j)$
by a suitably chosen {\em site} phase variables $\phi (i)$.
The natural choice for the $\phi (i)$'s is a simple average:
\begin{equation}
e^{i\phi (i)} = \frac{1}{4}\sum_{\sigma}e^{i\varphi (i,i+\sigma )}~~~,
\label{aii}
\end{equation}
where the sum over $\sigma$ runs over four bonds containing the site $i$.
With this choice of  $\phi (i)$'s we can replace:
\begin{equation}
e^{i\varphi (i,j)}\to e^{i[\phi (i) + \phi (j)]/2}~.
\label{aiii}
\end{equation}
Note that, given the choice of site variables (\ref{aii}),  
Eq. (\ref{aiii}) is an approximation, accurate up to second order
lattice derivatives of $\phi (i)$'s. We could use a more elaborate
representation of $\varphi (i,j)$'s in terms of $\phi (i)$'s
so that (\ref{aiii}) is satisfied to even higher degree of accuracy.
This, however, is entirely unnecessary in the present context, since
our overall accuracy is precisely at the level represented by
(\ref{aii}) and (\ref{aiii}). The replacement (\ref{aiii}) 
simplifies the pairing term 
of the lattice BdG Hamiltonian
and reproduces the form of the pairing operator
$\hat\Delta$ (\ref{IIiii}) used in our lattice
Hamiltonian of Eq. (\ref{IIi}). 

The above replacement of bond phases $\varphi (i,j)$'s with
site phases $\phi (i)$'s is also a necessary first step in our derivation
of the continuum BdG Hamiltonian. Since both $u(i)$ and $v(i)$ 
appearing in (\ref{ai}) are site fields we expect that the 
continuum pairing term in unconventional
superconductors will involve $u({\bf r})$ and 
$v({\bf r})$ acted upon by some {\em local}
operator. To determine the explicit form of 
this operator we first combine
(\ref{ai}) and (\ref{aii}) into:
\begin{eqnarray}
\frac{1}{2}\sum_i e^{i\phi (i)}\sum_{\sigma}D(i,i+\sigma)
e^{i\varphi (i,i+\sigma) - i\phi (i)} \nonumber \\
\times
\bigl [u^*(i)v(i+\sigma)+
u^*(i+\sigma)v(i)\bigr ] + {\rm h.c.}~~,
\label{aiv}
\end{eqnarray}
where we have transformed the summation 
over bonds into the summation
over sites. We now use $\varphi (i,i+\sigma) = \frac{1}{2}\phi (i)+
 \frac{1}{2}\phi (i+\sigma)+{\cal O}(\delta^2\phi)$,(\ref{aiii}) 
where $\delta^2$ denotes second order lattice derivatives which 
are unimportant in the continuum
limit. Also, from now on we restrict our 
attention to the most interesting case, a 
pure $d_{x^2-y^2}$ superconductor. This 
allows us to rewrite (\ref{aiv}) as:
$$\frac{1}{2}\sum_i \Delta (i)\bigl [-
e^{i\phi (i+{\bf\hat x})/2 - i\phi (i)/2}\bigl (u^*(i)v(i+{\bf\hat x})+
u^*(i+{\bf\hat x})v(i)\bigr ) $$
$$-e^{i\phi (i-{\bf\hat x})/2 - i\phi (i)/2}\bigl (u^*(i)v(i-{\bf\hat x})+
u^*(i-{\bf\hat x})v(i)\bigr )$$
$$+e^{i\phi (i+{\bf\hat y})/2 - i\phi (i)/2}\bigl (u^*(i)v(i+{\bf\hat y})+
u^*(i+{\bf\hat y})v(i)\bigr )$$
$$+e^{i\phi (i-{\bf\hat y})/2 - i\phi (i)/2}\bigl (u^*(i)v(i-{\bf\hat y})+
u^*(i-{\bf\hat y})v(i)\bigr )\bigr ]$$
\begin{equation}
+ {\rm h.c.},
\label{av}
\end{equation}
where $\Delta (i)\equiv\Delta\exp (i\phi(i))$ and ${\bf\hat x},{\bf\hat y}$ 
are unit displacements on the square lattice. Next, we expand
\begin{eqnarray}
e^{i\phi (i\pm {\bf\hat x}({\bf\hat y}))/2 - i\phi (i)/2} &\approx &
1 + \frac{i}{2}\bigl [\phi (i\pm {\bf\hat x}({\bf\hat y}))-\phi (i)\bigr ] 
\nonumber \\
&-&\frac{1}{8}\bigl [\phi (i\pm {\bf\hat x}({\bf\hat y}))-\phi (i)\bigr ]^2 
+\cdots ,
\end{eqnarray}
make a transition to continuum variables 
$u(i)\to au({\bf r})$, $v(i)\to av({\bf r})$,
$\Delta (i)\to\Delta ({\bf r})$, $\phi(i)\to\phi({\bf r})$, $
\sum_i\to\int(d^2r/a^2)$
and use the standard definitions of lattice derivatives to finally obtain
the continuum version of the pairing term:
$$
\frac{-a^2}{2}\int d^2r\bigl \{
u^*({\bf r})\Delta ({\bf r})\bigl [\bigl (\partial_x +\frac{i}{2}(
\partial_x\phi ({\bf r})\bigr ) \bigl (\partial_x +\frac{i}{2}(\partial_x\phi 
({\bf r})\bigr )v({\bf r})\bigr ]$$
$$
+\bigl [\bigl (\partial_x +\frac{i}{2}(\partial_x\phi ({\bf r})\bigr )
\bigl (\partial_x +\frac{i}{2}(\partial_x\phi ({\bf r})\bigr )u^*({\bf r})\bigr ] \Delta ({\bf r})v({\bf r})$$
\begin{equation}
 - (x\to y)\bigr\} + {\rm h.c.}~.
\label{avi}
\end{equation}
In going from (\ref{av}) to (\ref{avi}) one encounters some lengthy
but straightforward algebra. 
We found that decomposing the sum over
nearest neighbors in (\ref{av}) into
$s$-, $p$-, and $d$-wave components relative to site $i$ facilitates
the bookkeeping and makes the computations rather efficient.
All the relevant derivatives up to and including
second order are kept and accounted for. 
Higher order derivatives do not appear reflecting 
our original starting point of the nearest neighbor pairing only (\ref{ai}).
Note that $a$ is the lattice spacing in our model.

The form of the local continuum pairing operator is now apparent. We can view
$\Delta ({\bf r})=\Delta\exp (i\phi ({\bf r}))$ as representing the
{\em center-of-mass} portion of the gap function. The original non-locality,
arising from the {\em relative} $d_{x^2-y^2}$ character of the pairing, 
manifests itself through ``covariant" derivatives
$\partial_{\bf r} +\frac{i}{2}(\partial_{\bf r }\phi ({\bf r}))$, where
$\phi ({\bf r})$ is precisely the phase of $\Delta ({\bf r})$.
Note that (\ref{avi}) is explicitly invariant under the continuum version
of local gauge transformations:
$u({\bf r})\to u({\bf r})\exp(i\chi({\bf r}))$,
$v({\bf r})\to v({\bf r})\exp(-i\chi({\bf r}))$,
$\Delta ({\bf r})\to \Delta ({\bf r})\exp(2i\chi({\bf r}))$.

The off-diagonal elements of the Hamiltonian matrix appearing in 
the continuum BdG equations are obtained by taking the
functional derivatives of (\ref{avi}) with respect to $u^*({\bf r})$
and  $v({\bf r})$. This results in: 
\begin{eqnarray}
-a^2\{\partial_x,\{\partial_x,\Delta ({\bf r})\}\} +
a^2\{\partial_y,\{\partial_y,\Delta ({\bf r})\}\} - \nonumber \\
-\frac{i}{4}\Delta ({\bf r})a^2\bigl [(\partial_x^2\phi) - (\partial_y^2\phi)\bigr ],
\label{avii}
\end{eqnarray}
and its hermitian conjugate. Here we used the standard notation:
$\{\hat a,\hat b\}\equiv\frac{1}{2}(\hat a\hat b+\hat b\hat a)$. 
In performing the functional derivatives
we have exploited the fact that all spatial dependence of $\Delta ({\bf r})$
comes through its phase, i.e. $\partial_{\bf r}\Delta ({\bf r})
=i\Delta ({\bf r})\partial_{\bf r}\phi ({\bf r})$, in line with
our previous assumptions.

While our derivation starts with a familiar model of 
the lattice $d$-wave superconductor
(\ref{ai}) and naturally describes the $d_{x^2-y^2}$ state 
in actual continuum calculations it is often more convenient to 
consider a $d_{xy}$ superconductor, so that either an $x$ or a $y$ axis
coincide with a particular nodal direction, as in Section II.
We can obtain the pairing term in the continuum BdG Hamiltonian
of a $d_{xy}$ superconductor by simply rotating our result (\ref{avi})
by 45$^{\circ}$:
$$
\frac{-a^2}{2}\int d^2r\bigl\{
u^*({\bf r})\Delta ({\bf r})\bigl [\bigl (\partial_x +\frac{i}{2}(\partial_x\phi ({\bf r})
\bigr ) \bigl (\partial_y +\frac{i}{2}(\partial_y\phi ({\bf r})\bigr )
v({\bf r})\bigr ]$$
$$
+\bigl [\bigl (\partial_x +\frac{i}{2}(\partial_x\phi ({\bf r})\bigr )
\bigl (\partial_y +\frac{i}{2}(\partial_y\phi ({\bf r})\bigr )u^*({\bf r})\bigr ]\Delta ({\bf r})v({\bf r})$$
\begin{equation}
+ (x\to y)\bigr\} + {\rm h.c.}
\label{aviii}
\end{equation}

Similarly, by taking functional derivatives we obtain the off-diagonal
matrix elements of the continuum BdG Hamiltonian operator:
\begin{eqnarray}
-a^2\{\partial_x,\{\partial_y,\Delta ({\bf r})\}\} -
a^2\{\partial_y,\{\partial_x,\Delta ({\bf r})\}\} -\nonumber \\
-\frac{i}{2}\Delta ({\bf r})a^2\bigl (\partial_x\partial_y\phi\bigr )~~,
\label{avix}
\end{eqnarray}
and its hermitian conjugate, which is precisely the expression used in 
Section II, provided that we identify $p_F^{-2}$ with $2a^2$. 

The above derivation can be easily repeated for a $p$-wave lattice
Hamiltonian and is in fact only simpler. We therefore do not
give it explicitly but trust that the $d$-wave derivation provides
a sufficiently detailed prescription. Similarly, our derivation is
straightforwardly generalized to other unconventional forms
of superconducting pairing.


\section{Phase factors and superfluid velocities}

In this Appendix we derive expressions for superfluid velocities 
$\bv_s^A$ and $\bv_s^B$ which enter both continuum and lattice versions
of the BdG Hamiltonians in consideration in Section II. We start by taking 
the curl of Eq. (\ref{vsab}),
\begin{equation}
\nabla\times{\bf v}_s^\mu={2\pi\hbar\over m}\left[\hat{z}\sum_i
\delta(\br-\br_i^\mu)-\bB/\phi_0\right],
\label{cvsab}
\end{equation}
where $\phi_0=hc/e$ is the flux quantum, $\bB=\nabla\times\bA$,
and we have used Eq.\ (\ref{del}). In the intermediate field regime the 
magnetic field distribution is to an excellent approximation described
by the conventional London equation \cite{deGennes89},
\begin{equation}
\bB-\lambda^2\nabla^2\bB={1\over 2}\phi_0\hat{z}\sum_i\delta(\br-\br_i),
\label{london}
\end{equation}
where $\lambda$ is the London penetration depth and the sum now runs over
all vortex positions. The London equation is easily solved by going over to 
the Fourier space, obtaining $\bB(\br)=(2\pi)^{-2}\int{d^2k}e^{i\bk\cdot\br}
\bB_\bk$ with
\begin{equation}
\bB_\bk={1\over 2}\phi_0\hat{z}
{\sum_i e^{-i\bk\cdot\br_i}\over 1+\lambda^2 k^2}.
\label{b}
\end{equation}
If we now Fourier transform Eq.\ (\ref{cvsab}) we obtain
\begin{equation}
i\bk\times{\bf v}_{s\bk}^\mu={2\pi\hbar\over m}\left[\hat{z}\sum_i
e^{-i\bk\cdot\br_i}-\bB_\bk/\phi_0\right].
\label{fcvsab}
\end{equation}
To solve for ${\bf v}_{s\bk}^\mu$ we take a vector product of both sides
with $i\bk$. After substituting for $\bB_\bk$ and some easy algebra we 
obtain
\begin{equation}
{\bf v}_s^A={2\pi\hbar\over m}\int{d^2k\over (2\pi)^2}
{i\bk\times\hat{z}\over k^2}\left(A_\bk-{1\over 2}{A_\bk+B_\bk\over
1+\lambda^2k^2}\right) e^{i\bk\cdot\br},
\label{vsa}
\end{equation}
and a similar expression for ${\bf v}_s^B$ with $A_\bk$ and $B_\bk$
interchanged. Here we have defined
$$
A_\bk=\sum_ie^{-i\bk\cdot\br_i^A}, \ \ \ B_\bk=\sum_ie^{-i\bk\cdot\br_i^B}.
$$
Eq. (\ref{vsa}) gives an explicit formula for ${\bf v}_s^\mu$ which
can be evaluated for arbitrary distribution of vortices. For strongly type-II
materials  in fields well above $H_{c1}$ Eq.\ (\ref{vsa}) may be simplified 
further by rewriting the expression in the brackets as
$$
A_\bk{\lambda^2k^2\over 1+\lambda^2k^2}-{1\over 2}{B_\bk-A_\bk\over
1+\lambda^2k^2},
$$
and noting that since $\lambda^2k^2\sim\lambda^2/d^2\gg 1$ ($d$ being 
intervortex distance), the second term can be safely neglected. We thus
obtain 
\begin{equation}
{\bf v}_s^\mu={2\pi\hbar\lambda^2\over m}\int{d^2k\over (2\pi)^2}
{i\bk\times\hat{z}\over 1+\lambda^2k^2}
\sum_ie^{i\bk\cdot(\br-\br_i^\mu)},
\label{vsabf}
\end{equation}
a formula  used in Ref.\ \cite{franz1} which  is valid for 
all practical purposes. Phase factors ${\cal V}$ and ${\cal A}$ entering
the lattice Hamiltonians of Section III may be obtained as simple line
integrals of Eq.\ (\ref{vsabf}).




\begin{references}
\bibitem[*]{} Present address: Department of Physics and Astronomy, 
University of British Columbia, Vancouver, BC, Canada V6T 1Z1.
\bibitem{harlingen} D. J. Van Harlingen, Rev. Mod. Phys. {\bf 67}, 515 (1995).
\bibitem{lee} P. A. Lee, Science {\bf 277}, 50 (1997).
\bibitem{atkinson} See e.g. W.A. Atkinson, P.J. Hirschfeld, A.H. MacDonald, 
K. Ziegler, cond-mat/0005487 and references therein. 
\bibitem{Volovikdw93} G. E. Volovik, Sov.\ Phys.\ JETP Lett. 
{\bf 58}, 469 (1993).
\bibitem{sauls}S. K. Yip and J. A. Sauls, \prl {\bf 69}, 2264 (1992).
\bibitem{Rice} T. M. Rice, Nature {\bf 396}, 627 (1998); 
K. Ishida, H. Mukuda, Y. Kitaoka, et. al. Nature {\bf 396}, 658 (1998).
\bibitem{Volovikpw99} G. E. Volovik, cond-mat/9709159; Sov.\ Phys.\ JETP Lett.
{\bf 70}, 609 (1999).
\bibitem{Matsumoto99} M. Matsumoto and M. Sigrist, J. Phys. Soc. Jpn., 
{\bf 68} (3), 724 (1999); Physica B {\bf 281}, 973 (2000).
\bibitem{franz98b} M. Franz and A. J. Millis, 
Phys.\ Rev.\ B {\bf 58}, 14572 (1998). 
\bibitem{kwon99} H.-J. Kwon and A. T. Dorsey, \prb {\bf 60}, 6438 (1999). 
\bibitem{balents} L. Balents, M. P. A. Fisher and C. Nayak, \prb {\bf 60},
1654 (1999).
\bibitem{franz1} M. Franz and Z. Te\v{s}anovi\'c, \prl{\bf 84}, 554 (2000).
\bibitem{Caroli64} C. Caroli, P.G. de Gennes, J. Matricon, 
Phys. Lett. {\bf 9}, 307 (1964). 
\bibitem{deGennes89} P.~G. de~Gennes, {\em Superconductivity of Metals and 
Alloys} (Addison-Wesley, Reading, MA, 1989).
\bibitem{franz0} M. Franz and Z. Te\v{s}anovi\'c, \prl{\bf 80}, 4763 (1998).
\bibitem{resende98} X. R. Resende and A. H. MacDonald, Bull. Am. Phys. Soc. 
{\bf 43}, 573, 1998.
\bibitem{wang1} Y. Wang and A. H. MacDonald, \prb {\bf 52}, R3876 (1995).
\bibitem{kita1} K. Yasui and T. Kita, \prl {\bf 83}, 4168 (1999).
\bibitem{gorkov} L. P. Gor'kov and J. R. Schrieffer, 
Phys. Rev. Lett. {\bf 80}, 3360 (1998).
\bibitem{anderson} P. W. Anderson, cond-mat/9812063.
\bibitem{Melnikov} A. S. Melnikov, J. Phys. Cond. Matt. {\bf 82}, 4703 (1999).
\bibitem{marinelli} L. Marinelli, B. I. Halperin and S. H. Simon, 
\prb {\bf 62}, 3488 (2000).
\bibitem{moroz} A. Moroz, Phys. Rev. A {\bf 53}, 669 (1996).
\bibitem{Kopnin96} N. B. Kopnin and G.\ E.\ Volovik, JETP Lett., {\bf 64}
690 (1996).
\bibitem{ye1} J. Ye, cond-mat/0003251.
\bibitem{zirn} A. Altland, B. D. Simons and M. R. Zirnbauer, cond-mat/0006362.
\bibitem{sl1} S. H. Simon and P. A. Lee, \prl {\bf 78}, 1548 (1997).
\bibitem{remark1} We are grateful to J. Ye and N. Read for alerting us to this
fact. We also thank N. Read for an in-depth discussion of the issue of
gauge invariance of effective Hamiltonians.
\bibitem{remark2} We note that with the correct form of the pairing operator
(\ref{delta1}) this result is now exact. In Ref.\ \cite{franz1}  
it was necessary to neglect spurious higher order derivative terms which 
appeared as a consequence of the lack of the gauge invariance. 
\bibitem{jain1} J. K. Jain, \prl {\bf 63}, 199 (1989). 
\bibitem{halperin1} B. I. Halperin, P. A. Lee and N. Read, \prb {\bf 47}, 7312
(1993).
\bibitem{nielsen} M. Nielsen and P. Hedeg{\aa}rd, Phys. Rev. B {\bf 51}, 
7679 (1995).
\bibitem{remark3} Marinelli {\em et al.}\cite{marinelli} argued based on 
symmetry that there can be no true lines of nodes in the Brillouin zone. 
Even so, our numerical investigation indicates that
the small gap that exists in the linearized model closes up rapidly as 
$\alpha_D$ increases and becomes numerically indistinguishable from zero 
above $\alpha_D\approx 18$. For all practical purposes one may therefore 
speak about ``lines of nodes" in this context. 

\end{references}
\end{document}